\newif\iftwocol
\twocoltrue 

\iftwocol
\documentclass[journal,twocolumn,final]{IEEEtran-v17}
\else
\documentclass[journal,onecolumn,draftcls,12pt]{IEEEtran-v17}
\fi

\newlength{\figwidth}
\usepackage{color}
\definecolor{links}{rgb}{0.7,0,0}   
\definecolor{urls}{rgb}{0,0,0.8}    
\definecolor{cites}{rgb}{0,0,0.8}   

\usepackage[nosort]{cite}
\usepackage[intlimits]{amsmath}
\usepackage{bbm}
\usepackage{graphicx}
\usepackage{paralist}
\usepackage{fancyref}
\usepackage[stretch=16,shrink=16,step=4]{microtype}
\let\labelindent\relax 
\usepackage[inline]{enumitem}

\usepackage{header}

\usepackage{kolor}

\usepackage{kontext}

\usepackage{konfig}

\let\marginnote\undefined

\newcommand{\marginnote}[1]{\marginpar{\framebox{\framebox{#1}}}}

\newcommand{\newtext}[1]{#1}

\begin{document}

\title{Reliable Transmission of Short Packets through Queues and Noisy Channels under Latency and Peak-Age Violation Guarantees}
\author{
Rahul~Devassy,~\IEEEmembership{Student Member,~IEEE},
Giuseppe~Durisi,~\IEEEmembership{Senior Member,~IEEE},
Guido~Carlo~Ferrante,~\IEEEmembership{Member,~IEEE},
Osvaldo~Simeone,~\IEEEmembership{Fellow,~IEEE},
Elif~Uysal,~\IEEEmembership{Senior Member,~IEEE}
\thanks{R.\ Devassy and G. Durisi  are with the Department of Electrical Engineering, Chalmers University of Technology, Gothenburg, Sweden (e-mail: \{devassy,durisi\}@chalmers.se); G. C. Ferrante was with Chalmers University of Technology, Gothenburg, Sweden (e-mail: gcf@ieee.org); O.~Simeone is with King's College London, London, United Kingdom (e-mail: osvaldo.simeone@kcl.ac.uk); E.~Uysal is with Middle East Technical University, Ankara, Turkey (e-mail: uelif@metu.edu.tr).}
\thanks{Parts of this work were  presented at the IEEE International Symposium on Information Theory (ISIT), Jun. 2018~\cite{devassy18-06a}.}
\thanks{This work was partly funded by the Swedish Research Council under grant 2016-03293.
The simulations were performed in part on resources provided by the Swedish National Infrastructure for Computing (SNIC)  at C3SE.
O. Simeone has received funding from the European Research Council (ERC) under the European Union’s Horizon 2020 Research and Innovation Programme (Grant Agreement No. 725731).
E.~Uysal has received funding from T\"{U}B\.{I}TAK (Grant No. 117E215).}
}
%
%
\maketitle
\begin{abstract}
This work investigates the probability that the delay and the peak-age of information exceed a desired threshold in a point-to-point communication system with short information packets.
The packets are generated according to a stationary memoryless Bernoulli process,  placed in a single-server queue and then transmitted over a wireless channel.
A variable-length stop-feedback coding scheme---a general strategy that encompasses simple automatic repetition request (ARQ) and more sophisticated hybrid ARQ techniques as special cases---is used by the transmitter to convey the information packets to the receiver.
By leveraging finite-blocklength results,  the delay violation and the peak-age violation probabilities are characterized without resorting to approximations based on large-deviation theory as in previous literature.
Numerical results illuminate the dependence of delay and peak-age violation probability on system parameters such as the frame size and the undetected error probability, and on the chosen packet-management policy.
The guidelines provided by our analysis are particularly useful for the design of low-latency ultra-reliable communication systems.
\end{abstract}
\begin{IEEEkeywords}
Ultra-reliable low-latency communications, ARQ, HARQ, delay violation probability, peak-age violation probability.
\end{IEEEkeywords}
\section{Introduction}\label{sec:introduction}
Emerging wireless applications, such as \newtext{automated transportation, industrial automation and control, and tactile internet,} require the availability of mission-critical links that are able to deliver short information packets within stringent latency and reliability constraints.
The requirements for ultra-reliable low-latency communications (URLLC) defined by the international telecommunication union (ITU)~\cite{itu-r15-09a} underpin the introduction of the URLLC service in the next-generation wireless cellular system (5G).

Considering the isolated transmission of a single packet, new nonasymptotic tools in information theory have been applied for the design of URLLC since they can capture constraints on very high reliability, short coded packets, as well as short information payloads and sporadic transmissions.
Specifically, as recently shown in~\cite{polyanskiy2010channel,polyanskiy2011feedback}, finite-blocklength information theory provides accurate tools for describing the trade-offs among latency, reliability, and rate when transmitting individual short packets.

Communication latency is, however, not only determined by the blocklength, but also by the contribution of the queuing delays accrued in the presence of a data stream.
\newtext{Queuing delays are inherently random and designs that only control the average latency are not suitable to capture the stringent performance requirements of mission-critical applications.
Instead, solutions for URLLC should ensure that the overall delay is below a tolerable threshold with a sufficiently large probability.
Given the critical role of queuing delays in guaranteeing URLLC latency performance, we provide in this paper a joint coding-queuing analysis of the probability that the overall steady-state delay, including both queuing and transmission,  exceeds a desired level for a given reliability constraint.}

\newtext{As discussed, the overall latency is a key metric of interest in many use cases of URLLC. Latency is measured at the link layer and provides a useful quality-of-service measure for higher layers of the protocol. However, in some important applications, minimizing the delivery latency may not align with the requirements of the application layer. For example, in factory automation, the information packets exchanged over the wireless medium may carry sensor data needed to track a remote process at a given destination.
The primary performance metric at the application layer in such a scenario is not the delay of each packet, but rather the freshness of the sensor data available at the destination.
Packets that contain outdated sensor data are not valuable to the destination, and insisting on transmitting them with low latency is generally suboptimal.
A more relevant performance metric is the \textit{peak age of information}, which measures the maximum elapsed time since the last received update at the destination (see, e.g.,~\cite{costa2016age} and references therein).
To address this scenario, in this paper we also investigate the probability that the steady-state peak age of the information packets exceeds a desired level for a given reliability constraint.
}

\begin{figure}
\centering
\iftwocol
\includegraphics[width=0.95\columnwidth]{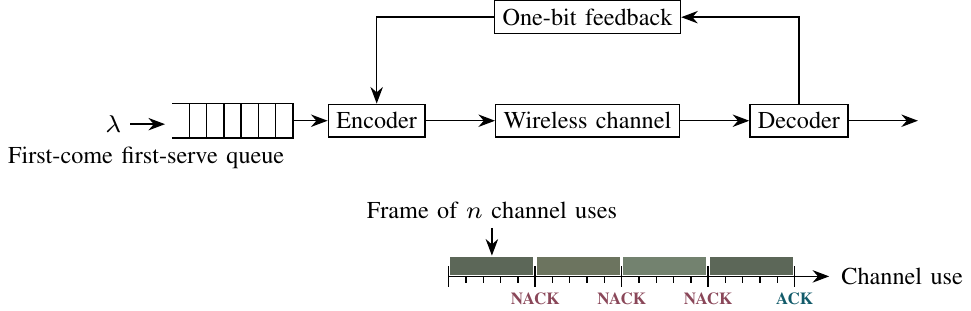}
\else
\includegraphics[width=0.6\columnwidth]{systemmodel}
\fi
\caption{In the considered system model, a new information packet arrives in each channel use with probability~$\packetarrivalprob$. The packet enters a first-come first-serve queue and it is then transmitted over a wireless channel using a variable-length stop-feedback code.}
\label{fig:system_model}
\end{figure}

Throughout the paper, we consider the
point-to-point communication system with random information-packet arrivals per channel use depicted in  Fig.~\ref{fig:system_model}.
The analysis assumes a single-server queue and the use of general variable-length-stop-feedback (VLSF) codes for information transmission~\cite[Eq.~(10)]{polyanskiy2011feedback}.
In a VLSF coding scheme, each codeword, which consists of an arbitrary large number of coded symbols,  is divided into frames of $n$ symbols.
After receiving a frame, the decoder attempts to recover the information packet.
Then, it communicates the outcome of the decoding attempt to the encoder through an ACK/NACK bit transmitted over a feedback channel.
If decoding fails, the next block is transmitted.
If decoding succeeds, transmission is stopped and the information packet is removed from the queue.
VLSF codes include as special cases strategies commonly employed in current wireless systems such as simple automatic repetition request (ARQ), where all frames corresponding to the same information packet contain the same coded bits, and incremental-redundancy hybrid ARQ (HARQ), where instead each new frame contains additional parity-check symbols.

\paragraph*{Related Work}
Aside from the work by Telatar and Gallager~\cite{telatar1995combining}, who employed an error-exponent approach, most  queuing analyses of communication links rely on a bit-pipe abstraction of the physical layer.
Accordingly, bits are delivered reliably at a rate equal to the channel capacity, or in the case of quasi-static fading channels, at a rate equal to the outage capacity for a given outage probability.
These works may be classified into three broad categories:
\begin{enumerate*}[label=(\roman*)]
\item analyses of the steady-state average delay;
\item \newtext{analyses of the delay violation probability using Chernoff bound via network calculus or large-deviation theory} (see~\cite{yeh12-a,al2016network} and references therein); and
\item analyses of the throughput-delay tradeoff under deadline constraints~\cite{zafer2009minimum,singh2015decentralized}.
\end{enumerate*}
However, the bit-pipe abstraction is not suitable for the analysis of URLLC traffic, given that the latency constraints prevent the use of channel codes with long blocklength.
Indeed, outage and ergodic capacity are poor performance benchmarks when packets are short~\cite{durisi16-09a}, and using them may result in inaccurate delay estimates.

Recognizing these limitations, Hamidi-Sepehr \emph{et al.}~\cite{hamidi2015delay} analyzed the queuing behavior when BCH codes are used.
They evaluated both the probability distribution of the steady-state queue size and the average delay.
Their analysis accounts for undetected error events, i.e., the event that an ACK is fed back although the decoded message is wrong.
Specifically, the authors showed how to mitigate the negative impact of such events by using the erasure decoding rule proposed in~\cite{ForneyExponentialBounds1968}.

A different approach, which relies on random coding, but is limited to the analysis of ARQ systems, is to replace capacity or outage capacity with the more accurate second-order fixed-blocklength approximations obtained in~\cite{polyanskiy2010channel,yang14-07c}.
This approach has been used in~\cite{Gursoy2013} to study the throughput achievable over a fading channel under a constraint on the probability of buffer overflow; \newtext{in~\cite{schiessl2018delay} to analyze the packet delay violation probability in the presence of imperfect channel-state information at the transmitter}; and in~\cite{she16-12b} to design the downlink of an ultra-reliable transmission system under a constraint on the end-to-end delay. 
\newtext{All these works rely on the characterization of tail probabilities using Chernoff bound via effective capacity~\cite{wu2003effective}, stochastic network calculus~\cite{jiang2008stochastic}, and effective bandwidth~\cite{chang1995effective}, and provide tight delay estimates only for large delay.}
Furthermore,  these results rely on the simplifying assumption that errors are perfectly detected at the decoder, i.e., the undetected error probability is zero.

An effective-capacity  analysis of general HARQ systems is given in~\cite{larsson2016effective}.
However, this analysis relies on an asymptotic information-theoretic approach based on   the renewal-reward theorem (see~\cite{caire01-07a} for details), which is again not suitable for URLLC, since it requires large packets.

A nonasymptotic random-coding lower bound on the rate achievable with VLSF codes for a given error probability and a given average blocklength has been recently proposed in~\cite[Thm.~3]{polyanskiy2011feedback}. As we shall see, this bound will turn out central to our analysis of general HARQ systems.

As mentioned, for applications in which packets carry status updates, the maximum time elapsed since the latest update available at the destination was generated at the source, commonly referred to as peak age of information, is more relevant than delay.
Most previous analyses of the peak age of information rely on simple physical-layer models.
A recent exception is~\cite{inoue2017stationary}, where the stationary distribution of the peak age is characterized, and~\cite{sun2017update}, where generalized age penalty functions are analyzed.
In these works, only an abstract model for the service process is considered.
For the special case of the binary erasure channel, where the undetected error probability is zero, analyses of the age for HARQ systems have been recently reported in~\cite{najm2017status,sac18-06a}.

\paragraph*{Contributions}
In this work, we analyze the delay and the peak-age violation probabilities achievable over a wireless channel in which information packets arrive in each channel use according to a discrete-time stationary memoryless Bernoulli process with parameter $\packetarrivalprob$.
We assume that the packets are transmitted  using a general VLSF coding scheme including ARQ and incremental-redundancy HARQ as special cases.
Our specific contributions are as follows:
\begin{itemize}[leftmargin=*]
\item Under first-come-first-serve (FCFS) queuing policy, which is optimal in terms of latency, we provide a novel definition of the steady-state delay violation probability that accounts not only for the event that the queuing plus the service time of a packet exceeds a given latency, but also for the undetected error events.
The sum of the complementary cumulative distribution function (CCDF) of the steady-state delay at the desired maximum latency and the undetected error probability is then used as a numerically computable upper bound on the steady-state delay violation probability.
\item For a fixed VLSF code, characterized by random stopping time $\servicestoppingtime$, frame size $\framesize$, and undetected error probability $\packeterrorprob$, we determine in closed form the probability generating function (PGF) of the steady-state delay as a function of the packet arrival rate $\packetarrivalprob$ and of the PGF of the stopping time $\servicestoppingtime$.
The steady-state delay CCDF is then computed numerically through an efficient inverse transform.
We also show how to accurately approximate this  transform using saddlepoint methods.

\item For the case of simple ARQ with perfect error detection, we illustrate how to evaluate the delay violation probability  using random-coding bounds from finite-blocklength information theory.
Specifically, we use the random-coding union bound with parameter $s$ (RCUs)~\cite[Thm.~1]{martinez11-02a}.
The resulting expression allows us to determine numerically the frame size $\framesize$  that minimizes the  delay violation probability for a given packet arrival rate $\packetarrivalprob$.
Our analysis reveals that an optimization of the frame size based only on average delay fails to minimize the delay violation probability.
In particular, considering a binary-input AWGN channel (bi-AWGN), we show that there exist two frame-size values resulting in the same average delay but yielding delay violation probabilities that differ by two orders of magnitude.
Finally, we show that the delay violation probability estimates based on the large deviation tools used in~\cite{schiessl2015delay} are loose, especially at low SNR.

\item For the more general case of HARQ, we adapt the random-coding VLSF achievability bound in~\cite[Thm.~3]{polyanskiy2011feedback}, which relies on threshold decoding, to obtain an estimate of both the PGF of the stopping time $\servicestoppingtime$ and the corresponding undetected error probability $\packeterrorprob$.
We show how to optimize the decoding threshold to trade optimally between the reduction of the steady-state delay CCDF at the desired maximum latency and the reduction of the undetected error probability, so as to minimize the overall delay violation probability.

\item We extend the analysis above to the scenario where one is interested in minimizing the peak-age violation probability, rather than the delay violation probability.
Focusing on the case of ARQ with perfect packet error detection, we study the impact  of the three different queue management policies considered in~\cite{costa2016age} on the peak age of information violation probability.
\newtext{These policies are: (i) FCFS with system capacity 1 (\DWTname), where information packets arriving while a packet is served are discarded}; \newtext{(ii) FCFS with system capacity 2 (\KTNname), where the system contains only the packet that is currently served and the first one arrived afterwards; and (iii) last-come first-serve with preemption in queue  (\KTLname)}, \newtext{where the system contains only the packet that is currently served and the last arrived packet.}
Our analysis supports the observation reported in~\cite{costa2016age} that the \newtext{\KTLname} policy outperforms the other two.
\newtext{Additionally, we also analyze a policy that is more suited to URLLC, in which whenever a new packet arrives, the transmission of the current packet is interrupted and the transmission of the new packet is started in the next frame.
This policy, which we refer to as LCFS with preemption in service (\ATLname), turns out to  outperform the other three policies in terms of peak age violation probability.}
\end{itemize}
The analysis in this paper extends the one reported in~\cite{devassy18-06a}, which was focused only on ARQ with perfect error detection and queues with FCFS  discipline.
\paragraph*{Notation}
Uppercase sans-serif letters denote random quantities and lightface letters denote deterministic quantities.
The distribution of a random variable~$\chinp$ is denoted by~$P_\chinp.$
With~$\Bexpectation{\cdot}$ we denote the expectation operator.
The indicator function and the ceiling function are denoted by~$\indicator{\cdot}$ and~$\ceil{\cdot}$, respectively.
We let~$\Bbernoullidist{p}$ denote a Bernoulli-distributed random variable with parameter~$p,$~$\Bbinomialdist{n}{p}$ a Binomial-distributed random variable with parameters~$n$ and~$p,$  and~$\Bgeometricdist{p}$ a geometrically distributed random variable with parameter~$p.$
The PGF of a nonnegative integer-valued random variable~$\chinp$ is $G_{\chinp}(s)= \Bexpectation{s^\chinp}$.
We use $\Rchinp^n$ to denote the vector $\parantheses{\Rchinp_1,\Rchinp_{2},\hdots,\Rchinp_n}$ where $n\geq 1$; finally, for a real number $y$, we let~$[y]^+=\max(0,y)$.

\section{System Model}\label{sec:system_model}

\newtext{We consider a general point-to-point discrete-time stationary memoryless channel with probability law}\footnote{\newtext{Extensions of the system model to channels with memory, which are relevant in the presence of fading, are discussed in Section~\ref{sec:fading_interference}}.}  $P_{\chout\given\chinp}$.
%
%
Assume that  a new information packet arrives in a given channel use with probability $\packetarrivalprob$ and that packet arrivals are independent across channel uses.
As a result,  the average interarrival time between information packets is $1/\packetarrivalprob$.
Upon its arrival, each packet, which is assumed to carry $k$ information bits, is stored at the transmitter in a single-server queue operating according to the FCFS policy.\footnote{We will consider more general packet-management policies in Section~\ref{sec:age}.}

An information packet that is ready to be transmitted over the channel is mapped into a coded packet by means of a VLSF encoder.
Specifically, each of the $2^k$ possible messages carried by an information packet is assigned to a distinct codeword of infinite length.
Then, the first frame, consisting of  the initial $\framesize$ symbols of the codeword associated to the desired message, is transmitted.
The VLSF decoder uses a stopping rule to decide whether the corresponding $n$ received symbols are sufficient to decode the message.
If they are not sufficient, the decoder sends a NACK message to the encoder via the feedback link, after which the encoder sends the next frame of~$\framesize$ symbols of the codeword.
This procedure continues until the stopping rule is triggered and the decoder produces an estimate of the transmitted message.
Finally, the decoder sends an ACK message to the encoder, which removes the packet from the queue and transmits the next packet in the queue.

We shall denote by $\servicestoppingtime$ 
the random number of frames needed for the transmission of a packet according to the described VLSF scheme.
Hence, the packet service time measured in channel uses is~$n\servicestoppingtime$.
Importantly, a packet may not necessarily be delivered correctly at the end of the service time, i.e., when an ACK is issued.
This occurs whenever the decoding process results in a wrong message estimate.
We denote the probability of this event, which corresponds to an undetected error, as $\packeterrorprob$.
Throughout, a VLSF coding scheme described as above is referred to as a $\parantheses{\framesize,\infobits,\probservicestoppingtime,\packeterrorprob}$--VLSF code where $\probservicestoppingtime$ is the probability distribution of $\servicestoppingtime$.

We start by considering a \textit{frame-synchronous} system, where time is organized into time frames of $n$ channel uses and the transmission of a codeword starts only at the beginning of a frame.
Under this assumption, if an information packet arrives when the buffer is empty, its transmission is delayed to the next available frame.
In Section~\ref{sec:relaxing_the_frame_synchronism}, we shall then relax this assumption and allow transmission to start in the next available channel use when the buffer is empty.
We refer to this alternative setup as  \emph{frame asynchronous}.
This alternative model yields a reduction in latency at the cost of a more involved frame-synchronization procedure.
In this paper, we shall assume synchronization to be ideal.

Under the frame-synchronous assumption, the system can be modeled as a $\mathit{Geo}/G/1$ queue with bulk arrivals, which is sometimes denoted $\mathit{Geo}^{[X]}/G/1$ (see \cite[Sec. 4.6.2]{bose2013introduction}).
This queue evolves along  the time index $t$ running over the time frames.
We next elaborate on this by detailing the arrival and departure processes.
All packets arriving within a time frame constitute a \emph{bulk}.
Let $\bulkarrivalat{t}$ be the number of packets received in the $t$-th time frame.
It follows that the bulk-arrival process $\curlybrac{\bulkarrivalat{t}}_{t=1}^{\infty}$ is stationary memoryless with $\Bbinomialdist{\framesize}{\packetarrivalprob}$ marginal distribution.
When $\bulkarrivalat{t}>0$, we say that a bulk has arrived at time frame $t$.
The arrival time $\arrivaltimeofbulk{m}$ of the $m$th bulk is, hence, defined as $\arrivaltimeofbulk{m} = \min\curlybrac{u:u>0,\sum_{t=1}^{u} \indicator{\bulkarrivalat{t}>0}=m}$.
The number of packets in the $m$th bulk is $\bulkarrivalcountat{m}=\bulkarrivalat{\arrivaltimeofbulk{m}}$.
We denote by~$\waitingtimeofbulk{m}$ the waiting time of the~$m$th bulk, i.e.,  the number of frames the first packet in the bulk remains in the queue before being served.
The service time of the~$m$th bulk, i.e., the total number of frames needed to transmit all packets in the~$m$th bulk, is denoted by $\servicetimeofbulk{m}$.
The service process $\curlybrac{\servicetimeofbulk{m}}_{m=1}^{\infty}$ is stationary memoryless with 
\begin{IEEEeqnarray}{rCl}
\servicetimeofbulk{m} &\sim& \sum_{k=1}^{\bulkarrivalcountat{m}}\servicetimeofpacket{k}\label{def:servicetime_bulk}
\end{IEEEeqnarray}
where the variables $\curlybrac{\servicetimeofpacket{k}}_{k=1}^{\bulkarrivalcountat{m}}$ are \iid  and independent of~$\bulkarrivalcountat{m}$, and
each~$\servicetimeofpacket{k}$ \newtext{is the random variable representing the} number of time frames needed to transmit one packet.
Since each packet is served using the same~$\parantheses{\framesize,\infobits,\probservicestoppingtime,\packeterrorprob}$--VLSF code, we have that $\servicetimeofpacket{k}$ has probability distribution $\probservicestoppingtime$.
%

We denote by $\queuesizeattime{t}$ the number of bulks remaining in queue at the start of the $(t+1)$th time frame.
Finally, the delay $\delayat{m}=\waitingtimeofbulk{m}+\servicetimeofbulk{m}$ of the $m$th bulk (measured in frames) is the sum of the waiting time $\waitingtimeofbulk{m}$ and the service time $\servicetimeofbulk{m}$.
For this queuing system, the process~$\curlybrac{\delayat{m}}_{m=1}^{\infty}$ admits a steady-state distribution as long as the \textit{traffic intensity} $\packetarrivalprob\framesize\Bexpectation{\servicestoppingtime}$, i.e., the ratio between average service time and average packet interarrival time, satisfies ~$\packetarrivalprob\framesize\Bexpectation{\servicestoppingtime}<1$~\cite[Thm. 11.3.5]{grimmett2001probability}.
\newtext{Indeed, if $\packetarrivalprob\framesize\Bexpectation{\servicestoppingtime}<1$, the Markov process describing the queue size is ergodic and, hence, it admits a stationary distribution.
This implies that~$\curlybrac{\delayat{m}}_{m=1}^{\infty}$ has a steady-state distribution, which is studied in the next section}.
The frame-asynchronous setup will be described in Section~\ref{sec:relaxing_the_frame_synchronism}.
\section{Steady-State Delay Violation Probability}\label{sec:steadystate_delay}
In this section, we shall define the steady-state delay violation probability and provide methods to compute it.
As detailed below, our definition accounts for both the probability that the delay of a packet exceeds a given latency constraint and for  undetected error events.
We start by formally defining an $\parantheses{\framesize,\infobits,\probservicestoppingtime,\packeterrorprob}$--VLSF code.
Then, we provide an upper bound on the steady-state delay violation probability that depends on the average packet interarrival rate $\packetarrivalprob$, on the PGF of the decoding time $\servicestoppingtime$, and on the undetected error probability $\packeterrorprob$ of the underlying VLSF code.
Finally, we relate $\packeterrorprob$ to the PGF of $\servicestoppingtime$ by deriving a variable-length random-coding bound based on the threshold decoding scheme proposed in~\cite[Thm.~3]{polyanskiy2011feedback}.
We shall also consider the special case of ARQ with perfect error detection, i.e., with $\packeterrorprob=0$.
For this case, we characterize the steady-state delay violation probability by leveraging the fixed-length nonasympotic information-theoretic bound on the smallest error probability achievable for a given frame length $\framesize$ and a given number of information bits $\infobits$ provided in~\cite[Thm.~1]{martinez11-02a}.

\subsection{Definition of a VLSF Code }\label{subsec:vlsf_definition}
The formal definition of a VLSF code provided next is similar to the one given in~\cite{polyanskiy2011feedback}, with the difference that, in our setup, decoding is attempted after each frame of $n$ channel uses, whereas in~\cite{polyanskiy2011feedback} decoding is attempted at every channel use.
Furthermore, we parametrize the code by using the probability distribution $\probservicestoppingtime$ of the random stopping time $\servicestoppingtime$, rather than its average $\Bexpectation{\servicestoppingtime}$.
\begin{defn}\label{def:vlsf_codes}
   An~$\parantheses{\framesize,\infobits,\probservicestoppingtime,\packeterrorprob}$--VLSF code consists of:
  \begin{enumerate}[leftmargin=*]
    \item A random variable~$\commonrand$, defined on a set~$\commonrandspace$,\footnote{We will discuss the cardinality of this set at the end of Section~\ref{subsec:delay_analysis_harq}.} whose realization is revealed to the encoder and the decoder before the start of transmission. The random variable~$\commonrand$ acts as common randomness and enables the use of randomized encoding and decoding strategies.
    \item \newtext{A randomized encoder~$f:\commonrandspace\times\messagespace\functionto\setX^{\infty}$, where $\messagespace=\{1,\dots,2^\infobits\}$, which maps the message~$\inpmessage \in \messagespace$ and the common randomness $\commonrand$ to an infinite-length codeword with symbols belonging to the set $\setX$.}
    We assume that~$\inpmessage$ is uniformly distributed on the set~$\messagespace$.
    \item \newtext{A sequence of decoders~$g_t:\commonrandspace\times\setY^{nt}\functionto\messagespace$, $t\geq 1$ that provide the estimate $\decoderoutput$ of the message~$\inpmessage$ at the~$t$th decoding attempt, on the basis of the $nt$ dimensional received signal vector~$\chout^{nt} \in \setY^{nt}$.}
    \item A nonnegative integer-valued random variable~$\stoppingtime$, which is a stopping time that depends only on the common randomness $\commonrand$ and on the received frames.
    \item The final estimate of~$\inpmessage$
    \begin{equation}\label{eq:decoded_msg}
        \decoderoutput = g_\stoppingtime\parantheses{\commonrand,\chout^{n\stoppingtime}}
    \end{equation}
     which satisfies the error-probability constraint $ \probof{\decoderoutput\neq\inpmessage} \leq \packeterrorprob$.
  \end{enumerate}
\end{defn}
\subsection{Definition of the Delay Violation Probability}\label{subsec:delay_metric_and_pgf}
For the practically relevant scenario considered in this paper in which the VLSF decoding process may yield undetected error events, it is not satisfactory to define the delay violation probability simply as the CCDF of the steady-state delay---the sum of service time $\servicestoppingtime$ and queuing time of a packet at steady state. This is because an undetected error event results in a packet being removed from the queue, although it has not been correctly delivered.
To overcome this problem, we define the delay violation probability so as to account for both the event of the delay exceeding a maximum latency  and the case of the message being decoded incorrectly.

Formally, for a given latency constraint $\delaythreshold$ measured in channel uses, we define  the steady-state delay violation probability $\Bprobofdelayviolationactual{\delaythreshold}$ \newtext{for the case  $\packetarrivalprob\framesize\Bexpectation{\servicestoppingtime}<1$} as
%
\begin{IEEEeqnarray}{rCl}
\IEEEeqnarraymulticol{3}{l}{\Bprobofdelayviolationactual{\delaythreshold}}\nonumber \\
 & = & \limsup_{m\to\infty} \biggl[ 1-\probof{\delayofpacket{m}<\delaythreshold/\framesize \text{ and } \messageofpacket{m}=\decodedmessageofpacket{m}}\biggr],\IEEEeqnarraynumspace \label{expr:delayviolation_criteria}
\end{IEEEeqnarray}
where~$\delayofpacket{m}$ denotes the delay of the~$m$th packet measured in number of frames, and $\messageofpacket{m}$ and $\decodedmessageofpacket{m}$ are the message corresponding to the~$m$th packet and its estimate~\eqref{eq:decoded_msg} at the decoder, respectively.\footnote{Note that we define the delay violation event in terms of packet delay rather than bulk delay.
This explains the presence of the superscript $(\mathrm{p})$ to the delay random variable $\delayofpacket{m}$.
This definition has the advantage to hold for both the frame-synchronous and the frame-asynchronous cases.}

Analyzing~\eqref{expr:delayviolation_criteria} directly seems to be prohibitive, apart from the special case $\packeterrorprob=0$, which will be discussed in Section~\ref{subsec:delay_analysis_arq}.
Therefore, we will focus in the \newtext{remainder} of the paper on the following upper bound $\Bprobofdelayviolation{\delaythreshold}$ on $\Bprobofdelayviolationactual{\delaythreshold}$, which is obtained by applying the union bound:
\begin{IEEEeqnarray}{rCl}
  \IEEEeqnarraymulticol{3}{l}{ \Bprobofdelayviolationactual{\delaythreshold}  } \nonumber \\
&\leq & \limsup_{m\to\infty}\biggl[ \probof{\delayofpacket{m}\geq\delaythreshold/\framesize}+\probof{ \messageofpacket{m}\neq\decodedmessageofpacket{m}}\biggr]\IEEEeqnarraynumspace\\
&\newtext{\leq} & \limsup_{m\to\infty} \probof{\delayofbulk{m}\geq\delaythreshold/\framesize}+\packeterrorprob\label{expr:delayviolation_criteria_bound_int_1}\\
&=& \probof{\delay\geq\delaythreshold/\framesize} + \packeterrorprob \label{expr:delayviolation_criteria_bound_int_2}\\
&\triangleq & \Bprobofdelayviolation{\delaythreshold}.\label{expr:delayviolation_criteria_bound}
\end{IEEEeqnarray}
%
Here,~\eqref{expr:delayviolation_criteria_bound_int_1} follows because the delay of a bulk coincides with the delay of the last packet in the bulk; and in~\eqref{expr:delayviolation_criteria_bound_int_2}, $\delay$ denotes a random variable whose distribution coincides with the steady-state distribution of $\delayofbulk{m}$.
The upper bound $\Bprobofdelayviolation{\delaythreshold}$ is the sum of the CCDF of the steady-state delay, computed at the latency constraint,  and the undetected error probability.
In the reminder of this section, we will discuss the computation of these two quantities.

Before doing so, however, it is appropriate to point out that other notions of delay violation probability are possible.
For example, if each message carries  mission-critical information, it may be preferable to avoid decoding the message, hence violating the latency constraint, rather than decoding it incorrectly.
In such a scenario, one may replace the sum in~\eqref{expr:delayviolation_criteria_bound_int_2} by another suitably chosen function (e.g., a weighted sum).
Our analysis can be readily extended to such a scenario.

We start by evaluating the CCDF of the steady-state delay.
As a first step, we provide in Theorem~\ref{thm:delay_sync_steadystate} below the PGF  of $\delay$ as a function of the  packet arrival rate $\packetarrivalprob$ and of the PGF  of the service time $\stoppingtime$.

\begin{thm}\label{thm:delay_sync_steadystate}
For a given packet interarrival rate $\packetarrivalprob$ and for every~$\parantheses{\framesize,\infobits,\probservicestoppingtime,\packeterrorprob}$--VLSF code satisfying the stability condition $\packetarrivalprob\framesize\Bexpectation{\servicestoppingtime}<1$, the PGF of the steady-state delay $\delay$ for the frame-synchronous model is
\iftwocol
\begin{IEEEeqnarray}{rCl}
 \BPGFof{\delay}{s} &=& \parantheses{1-\packetarrivalprob\framesize\Bexpectation{\servicestoppingtime}}\nonumber\\
 &&.\frac{\parantheses{1-s}\parantheses{\parantheses{1-\packetarrivalprob}^\framesize-\parantheses{1-\packetarrivalprob+\packetarrivalprob \BPGFof{\servicestoppingtime}{s}}^\framesize}} {\parantheses{1-\parantheses{1-\packetarrivalprob}^\framesize}\parantheses{s-\parantheses{1-\packetarrivalprob+\packetarrivalprob \BPGFof{\servicestoppingtime}{s}}^\framesize}}.\label{expr:pgf_steadystate_delay}
\end{IEEEeqnarray}
\else
\begin{IEEEeqnarray}{rCl}
 \BPGFof{\delay}{s} &=& \parantheses{1-\packetarrivalprob\framesize\Bexpectation{\servicestoppingtime}}     \frac{\parantheses{1-s}\parantheses{\parantheses{1-\packetarrivalprob}^\framesize-\parantheses{1-\packetarrivalprob+\packetarrivalprob \BPGFof{\servicestoppingtime}{s}}^\framesize}} {\parantheses{1-\parantheses{1-\packetarrivalprob}^\framesize}\parantheses{s-\parantheses{1-\packetarrivalprob+\packetarrivalprob \BPGFof{\servicestoppingtime}{s}}^\framesize}}.\label{expr:pgf_steadystate_delay}
\end{IEEEeqnarray}
\fi
\end{thm}
\begin{IEEEproof}
See Appendix~\ref{proof:delay_sync_steadystate}.
\end{IEEEproof}

The CCDF of  $\delay$ can be obtained from the PGF in~\eqref{expr:pgf_steadystate_delay} using the inversion formula
\begin{IEEEeqnarray}{rCl}
\probof{\delay\geq\delaythreshold/\framesize} &=&  1- \parantheses{\frac{\indicator{d\geq 2}}{2\pi i}\oint\nolimits_{\setC} \frac{\BPGFof{\delay}{s}}{(1-s)s^{d-1}}\infinitesimal s},
\IEEEeqnarraynumspace \label{expr:ccdf_steadystate_delay}
\end{IEEEeqnarray}
where~$d=\left\lceil\delaythreshold/\framesize\right\rceil$ and~$\setC$ is a circle centered at the origin enclosing all poles of $\BPGFof{\delay}{s}/(1-s).$
Since the contour integral in~\eqref{expr:ccdf_steadystate_delay} is not known in closed form, the numerical evaluations of~\eqref{expr:pgf_steadystate_delay} we shall present in Section~\ref{sec:numerical_results} are based on a recursion based $z$-transform inversion of %
$\BPGFof{\delay}{s}/(1-s)$.
This method, which is based on~\cite[Eq. (10)]{jenkins1967useful}, is detailed in Appendix~\ref{appendix:CCDF_to_PGF}.

A reduced-complexity approach to compute the delay violation probability from~\eqref{expr:pgf_steadystate_delay} is through the saddlepoint method~\cite[Eq. (2.2.10)]{jensen1995saddlepoint}.
Let $\Bexpectation{\delay}$ be the expectation of the steady-state delay.
Note that we can obtain  $\Bexpectation{\delay}$ from the PGF of $\delay$ in\newtext{~\eqref{expr:pgf_steadystate_delay}} by computing the limit~$\Bexpectation{\delay}=\lim_{s\uparrow 1}G_{\delay}'\parantheses{s}$
where we use the prime notation to denote derivatives.
To provide the saddle-point approximation, we distinguish between the two cases $\left\lceil\delaythreshold/\framesize\right\rceil> \Bexpectation{\delay}$
and
$\left\lceil\delaythreshold/\framesize\right\rceil< \Bexpectation{\delay}$.
In the first case, we have
\begin{IEEEeqnarray}{rCl}
\probof{\delay\geq\delaythreshold/\framesize} &\approx&  \frac{B_0\parantheses{\theta\sigma\parantheses{\theta}}}{\sigma\parantheses{\theta}\parantheses{1-e^{-\theta}}} e^{\kappa\parantheses{\theta}-\theta \left\lceil\delaythreshold/\framesize\right\rceil}, \label{expr:ccdf_steadystate_delay_saddlepoint}
\end{IEEEeqnarray}
where %
 $\kappa\parantheses{x} = \Blog{\BPGFof{\delay}{e^x}}$, %
 $\theta = \argmin_{x\in\mathbb{R}} \kappa\parantheses{x}-x\left\lceil\delaythreshold/\framesize\right\rceil$, %
 $\sigma\parantheses{x} =  \sqrt{\kappa''\parantheses{x}}$, and %
 $B_0\parantheses{x} = xe^{x^2/2}Q\parantheses{x}$, %
 where~$Q\parantheses{x}$ is the Gaussian Q-function.
In the second case, we use that $\probof{\delay\geq\delaythreshold/\framesize}=1-\probof{-\delay\geq-1-\left\lceil\delaythreshold/\framesize\right\rceil}$ and then apply~\eqref{expr:ccdf_steadystate_delay_saddlepoint} to the second term.
As we shall show in Section~\ref{sec:numerical_results}, the approximation~\eqref{expr:ccdf_steadystate_delay_saddlepoint} is very accurate over a large range of system parameters.

To evaluate $\Bprobofdelayviolation{\delaythreshold}$ using~\eqref{expr:ccdf_steadystate_delay} or~\eqref{expr:ccdf_steadystate_delay_saddlepoint}, one needs  the PGF of the decoding time $\stoppingtime$ and the undetected error probability $\packeterrorprob$.
Next, we discuss how to obtain these quantities.
We will start from the simpler case of ARQ with perfect error detection in Section~\ref{subsec:delay_analysis_arq}, and then move to the more general VLSF setup with positive undetected error probability in Section~\ref{subsec:delay_analysis_harq}.
\subsection{A Special Case: ARQ with Perfect Error Detection}\label{subsec:delay_analysis_arq}
In the ARQ setup, the VLSF encoder repeats the same $n$ channel inputs in each frame until it receives an ACK.
We will analyze this coding scheme under the commonly used simplifying assumption of perfect error detection, namely, an ACK is fed back only if the message is decoded correctly.
This can be achieved, for example, through the use of hashing methods, whose overhead we shall neglect.
We will consider the more general and practically relevant case of VLSF codes with positive undetected error probability in Section~\ref{subsec:delay_analysis_harq}.
Under the assumption of perfect error detection, we have $\packeterrorprob=0$ and, hence, $\Bprobofdelayviolation{\delaythreshold}=\Bprobofdelayviolationactual{\delaythreshold}$.

As we shall see shortly,  the PGF of $\stoppingtime$ is uniquely determined in this case by the frame error probability $\epsilon$ of the fixed-blocklength $(k,n)$ channel code that is used to map the information packet into the $n$ coded symbols transmitted within a frame.
The RCUs bound~\cite[Thm.~1]{martinez11-02a} provides the following random-coding bound on $\epsilon$, as a function of the code parameters $k$ and $n$.
\begin{thm}[\newtext{\cite[Thm.~1]{martinez11-02a}}]\label{thm:rcus}
Let $\chinp^n$ be an \iid $n$-dimensional vector with \newtext{\iid entries distributed according to $P_{\chinp}$.}
For a given $\alpha>0$,\footnote{This parameter is traditionally denoted as $s$, hence the name of the bound.
We denote it by $\alpha$ since we use $s$ to denote the argument of PGFs.}
let the generalized information density be defined as
\begin{equation}
    \infodensitys{\Rchinp_1^\framesize}{\Rchout_1^\framesize}=\sum_{k=1}^{\framesize}\log\frac{P_{\chout\given\chinp}\parantheses{\Rchout_k\given\Rchinp_k}^\alpha}{\Bexpectation{P_{\chout\given\chinp}\parantheses{\Rchout_k\given \chinp_k}^\alpha}}.
\end{equation}
Then, there exists a $(k,n)$-code with average frame error probability $\packeterrorarq$ satisfying
\begin{equation}
    \packeterrorarq\leq\packeterrorarqrcus\triangleq \inf_{\alpha\geq 0} \Bexpectation{\exp\curlybrac{-\Bigl[\infodensitys{\chinp^\framesize}{\chout^\framesize}-\log(2^k-1)\Bigr]^{+}}} \label{expr:rcus_biawgn}
\end{equation}
\newtext{where the expectation is with respect to the joint probability distribution
\begin{equation}\label{eq:prob_dist_rcus}
    P_{\chinp^n\chout^n}(\Rchinp^n,\Rchout^n)=\prod_{k=1}^n P_{\chinp}(\Rchinp_k)P_{\chout\given\chinp}(\Rchout_k\given\Rchinp_k).
\end{equation}
}%
\end{thm}
Using the achievable error probability $\packeterrorarqrcus$, we conclude that the decoding time is distributed as $\servicestoppingtime\sim\Bgeometricdist{1-\packeterrorarqrcus}$.
Hence, its PGF is~$\BPGFof{\servicestoppingtime}{s} = {\parantheses{1-\packeterrorarqrcus}s}/\parantheses{1-\packeterrorarqrcus s}$.
%

\paragraph*{An Alternative Bound Through a Large-Deviation Analysis}
A common approach in the queuing literature to  characterize the CCDF of the steady-state delay is \newtext{through the Chernoff bound}.
When the packet service time is deterministic and constant, this approach is the effective bandwidth method~\cite{guerin1991equivalent,chang1995effective}, whereas  when the arrival rate is deterministic and constant and the service rate is random one uses the effective capacity method~\cite{wu2003effective}.
In our model, both the arrival and the service processes are random.
Hence, these two tools cannot be used.
A suitable tool for our setup, under the additional assumption of ARQ with perfect error detection, is stochastic network calculus~\cite{jiang2008stochastic}.
In the next theorem, we present an upper bound on $\Bprobofdelayviolationactual{d_0}$ based on stochastic network calculus, which is obtained by adapting~\cite[Thm.~1]{schiessl2015delay} to our setup.\footnote{ \newtext{The proof of this result, which  follows closely the analysis in~\cite[Sec. 4.4]{schiessl2015delay}, is omitted due to space constraint.}}
As we shall see, this bound is sometimes much looser than the exact characterization of $\Bprobofdelayviolationactual{d_0}$ obtainable through~\eqref{expr:ccdf_steadystate_delay} or~\eqref{expr:ccdf_steadystate_delay_saddlepoint}.
\begin{thm}\label{thm:ccdf_steadystate_delay_netcalc}
For the ARQ setup and under the assumption of perfect error detection, the steady-state delay violation probability~$\Bprobofdelayviolationactual{\delaythreshold}$  is upper-bounded as
\begin{equation}
\Bprobofdelayviolationactual{\delaythreshold}  \leq \hspace{-2ex} \inf_{\substack{s>1:\\\BPGFof{\bulkarrivalat{}}{s}\BPGFof{\servicedbitsperslot}{1/s}<1}} \frac{\BPGFof{\servicedbitsperslot}{1/s}^{d-1}}{1-G_{\bulkarrivalat{}}\parantheses{s}\BPGFof{\servicedbitsperslot}{1/s}},\label{expr:ccdf_steadystate_delay_netcalc}
\end{equation}
where~$d=\left\lceil\delaythreshold/\framesize\right\rceil$ and the PGFs $\BPGFof{\bulkarrivalat{}}{s}$ and $\BPGFof{\servicedbitsperslot}{s}$ are~$\BPGFof{\bulkarrivalat{}}{s} = \parantheses{1-\packetarrivalprob+\packetarrivalprob s}^\framesize$ and $
\BPGFof{\servicedbitsperslot}{s} = \packeterrorarqrcus + \parantheses{1-\packeterrorarqrcus} s$.
\end{thm}
%
Note that the numerical evaluation of~\eqref{expr:ccdf_steadystate_delay_netcalc} has the same complexity as the evaluation of the saddle-point approximation~\eqref{expr:ccdf_steadystate_delay_saddlepoint}, which, as we shall see, is more accurate.
Both expressions are easier to evaluate than~\eqref{expr:ccdf_steadystate_delay}.
\subsection{VLSF Codes with Positive Undetected Error Probability}\label{subsec:delay_analysis_harq}
We present next random-coding characterizations of the CCDF of the stopping time $\stoppingtime$ of a $\parantheses{\framesize,\infobits,\probservicestoppingtime,\packeterrorprob}$--VLSF code, and of the corresponding undetected error probability $\packeterrorprob$, which are based on an adaptation of the threshold-decoding achievability bound provided in~\cite[Thm.~3]{polyanskiy2011feedback}.

\begin{thm}\label{thm:vlsf_achievability}
Fix a positive real number~$\infodensitythreshold$.
Let~$\chinp=\parantheses{\chinp_1,\chinp_2,\cdots}$ be a stationary memoryless stochastic process \newtext{with~$\chinp_1\distas P_{\chinp}$}.
Let~$\auxchinp$ be an independent copy of the same process. Define a sequence of information density functions~$\infodensity{\Rchinp^t}{\Rchout^t} = \sum_{k=1}^{t}\Blogfrac{ P_{\chout\given\chinp}\parantheses{\Rchout_k\given\Rchinp_k}}{ P_{\chout}\parantheses{\Rchout_k}}, \, t\geq 1,$
%
%
where $P_\chout(\cdot)$ is the output distribution induced by the distribution $P_{\chinp}$ on the input symbols through the channel law $P_{\chout\given\chinp}$.
Also define the  stopping times
\begin{IEEEeqnarray}{rCl}
\thresholdstoppingtime &=& \inf\curlybrac{t\geq 1 :\infodensity{\chinp^{nt}}{\chout^{nt}}\geq\infodensitythreshold} \label{eq:info_dens_threshold}\\
\auxthresholdstoppingtime &=& \inf\curlybrac{t\geq 1 :\infodensity{\auxchinp^{nt}}{\chout^{nt}}\geq\infodensitythreshold}.\label{eq:info_dens_threshold_aux}
\end{IEEEeqnarray}
Then, there exists an~$\parantheses{\framesize,\infobits,\probservicestoppingtime,\packeterrorprob}$--VLSF code with
\begin{equation}
    \probof{\servicestoppingtime\geq t} \leq \probof{\thresholdstoppingtime\geq t},\quad t\geq 0\label{expr:stoppingtime_bound}
\end{equation}
and
\begin{equation}
\packeterrorprob \leq \parantheses{2^\infobits-1}\probof{\auxthresholdstoppingtime\leq\thresholdstoppingtime}.\label{expr:errorprob_bound}
\end{equation}
\end{thm}
\begin{IEEEproof}
See Appendix~\ref{proof:vlsf_achievability}.
\end{IEEEproof}
In words, we can think of $\thresholdstoppingtime$ as the random variable describing the frame index in which the accumulated information density corresponding to the transmitted codeword exceeds the threshold~$\infodensitythreshold$.
Similarly, $\auxthresholdstoppingtime$ is the frame index in which the accumulated information density corresponding to a codeword different from the transmitted one exceeds~$\infodensitythreshold$.
Then, as formalized in~\eqref{expr:stoppingtime_bound}, $\thresholdstoppingtime$ stochastically dominates $\servicestoppingtime$.
Furthermore, as shown in~\eqref{expr:errorprob_bound}, we have an undetected error if the inequality $\auxthresholdstoppingtime\leq \thresholdstoppingtime$ holds for some incorrect codeword.

In Section~\ref{sec:numerical_results}, we will use Theorem~\ref{thm:vlsf_achievability} to evaluate the upper bound $\Bprobofdelayviolation{\delaythreshold}$ on the delay violation probability defined in~\eqref{expr:delayviolation_criteria_bound}.
Note that by replacing in~\eqref{expr:pgf_steadystate_delay} the actual stopping time   $\servicestoppingtime$  of the threshold-based VLSF code in Theorem~\ref{thm:vlsf_achievability}  with $\thresholdstoppingtime$, one obtains an upper bound on $\Bprobofdelayviolation{\delaythreshold}$ because of the stochastic dominance defined by~\eqref{expr:stoppingtime_bound}.

\paragraph*{Handling Truncated Hybrid ARQ} In Theorem~\ref{thm:vlsf_achievability}, we allow for an infinite number of retransmissions, which is not feasible in practice.
Our achievability bound can be easily adapted to include the requirement that the number of retransimission is limited---a scenario sometimes referred to as truncated hybrid ARQ.
The corresponding bound is given in the
following corollary, whose proof, which is omitted, is a simple adaptation of the proof of Theorem~\ref{thm:vlsf_achievability}.
\begin{cor}\label{thm:vlsf_achievability_finite_allowedreps}
Let~$\thresholdstoppingtime$ and $\auxthresholdstoppingtime$ be as defined in~\eqref{eq:info_dens_threshold} and~\eqref{eq:info_dens_threshold_aux}, respectively.
Then, there exist an~$\parantheses{\framesize,\infobits,\probservicestoppingtime,\packeterrorprob}$--VLSF code with codewords spanning at most $\allowedreps$ frames satisfying
\begin{equation}
  \probof{\servicestoppingtime\geq t} \leq \probof{\min\parantheses{\allowedreps,\thresholdstoppingtime}\geq t},\ 1\leq t\leq \allowedreps\label{expr:stoppingtime_bound_finite_allowedreps}
\end{equation}
and
\begin{equation}
 \packeterrorprob \leq \parantheses{2^\infobits-1}\probof{\auxthresholdstoppingtime\leq\min\parantheses{\allowedreps,\thresholdstoppingtime}}+\probof{\thresholdstoppingtime\geq \allowedreps+1}.\label{expr:errorprob_bound_finite_allowedreps}
\end{equation}
\end{cor}

The additional term on the right-hand side of~\eqref{expr:errorprob_bound_finite_allowedreps} compared  to~\eqref{expr:errorprob_bound} captures the case in which a NACK is sent after the transmission of the last frame, which yields a detected error.

As a side remark, we note that, for the case in which the number of frame transmissions does not exceed a maximum number~$\allowedreps$, the probability distribution of $\servicestoppingtime$ can be described by providing an $(\allowedreps-1)$-dimensional vector.
This implies that the set of pairs $(\probservicestoppingtime,\packeterrorprob)$ for which an $\parantheses{\framesize,\infobits,\probservicestoppingtime,\packeterrorprob}$--VLSF code exists can be mapped into a subset of $\reals^{\allowedreps}$.
It follows from Definition~\ref{def:vlsf_codes} that this subset is the convex hull of the set of points in $\reals^{\allowedreps}$ corresponding to deterministic (i.e., non-randomized) $\parantheses{\framesize,\infobits,\probservicestoppingtime,\packeterrorprob}$--VLSF codes.
Applying Carath\'eodory's theorem, we conclude that every $\parantheses{\framesize,\infobits,\probservicestoppingtime,\packeterrorprob}$--VLSF code can be written as a convex combination of $\allowedreps+1$ deterministic VLSF codes.
This implies that the common randomness $\commonrand$ in Definition~\ref{def:vlsf_codes} belongs to a set of cardinality no larger than $\allowedreps+1$.


Note that a stochastic-network calculus upper bound on $\Bprobofdelayviolation{\delaythreshold}$, similar to the one we reported in Theorem~\ref{thm:ccdf_steadystate_delay_netcalc} for the case of simple ARQ with perfect error detection, is not available in the more general VLSF setup considered here.
Indeed, a crucial assumption in the proof of Theorem~\ref{thm:ccdf_steadystate_delay_netcalc} is that the random process describing the number of packets leaving the queue in a given frame is stationary and memoryless.
This assumption does not hold in the general VLSF setup.
\section{The Frame-Asynchronous Setup} 
\label{sec:relaxing_the_frame_synchronism}
We next consider an alternative setup in which, if the buffer is empty when a packet arrives, the corresponding codeword is transmitted starting from the next available channel use.
We  refer to this model as  frame asynchronous.
This setup can be modeled as a simple~$Geo/G/1$ queue.
The PGF of the steady-state delay $\delayasync$ measured in channel uses is given in the following theorem.
\begin{thm}\label{thm:delay_async_steadystate}
For a given packet interarrival rate $\packetarrivalprob$ and for every~$\parantheses{\framesize,\infobits,\probservicestoppingtime,\packeterrorprob}$--VLSF code satisfying the stability condition $\packetarrivalprob\framesize\Bexpectation{\servicestoppingtime}<1$, the PGF of the steady-state delay for the frame-asynchronous model is $\BPGFof{\delayasync}{s} = \parantheses{1-\packetarrivalprob\framesize\Bexpectation{\servicestoppingtime}} \parantheses{s-1} \BPGFof{\servicestoppingtime}{s^n}/\parantheses{s-1+\packetarrivalprob \parantheses{1-\BPGFof{\servicestoppingtime}{s^n}}}$.
\end{thm}
\begin{IEEEproof}
After setting $\bulkarrivalcountat{1}=1$, $\bulkarrivalat{1}\sim\Bbernoullidist{\packetarrivalprob}$, and noting that $\servicetimeofbulk{1}$ has the same distribution as~$n\servicestoppingtime$,
one uses the same steps as in the proof of Theorem~\ref{thm:delay_sync_steadystate}.
\end{IEEEproof}
The CCDF of the steady-state delay and its saddlepoint approximation can be obtained  by proceeding as in Section~\ref{subsec:delay_metric_and_pgf}.
Also for this setup, a stochastic network calculus bound on the delay violation probability similar to Theorem~\ref{thm:ccdf_steadystate_delay_netcalc} is not feasible since the random process describing the number of packets leaving the queue at each channel use is not stationary memoryless.
\section{Steady-State Peak-Age Violation Probability}\label{sec:age} 
We next characterize the violation probability of the steady-state peak age at the application layer.
As in~\cite{costa2016age}, we assume that the destination is interested in timely updates about the status of a random process observed at the source.
Each of the information packets generated by the source contains a sample of this random process and the time at which the sample was taken.
We focus only on the frame-synchronous model, and assume that the time stamp of a packet is the frame index at which the packet enters the queue.
So the $m$th packet carries the time index $\arrivaltimeofpacket{m}$.
Since packets arriving in the same frame carry the same time stamp, we shall assume throughout this section that only one of these packets is allowed to enter the queue, whereas the other packets are discarded.
In the terminology we introduced in Section~\ref{sec:system_model}, this corresponds to assuming that each bulk of packets contains a single packet.

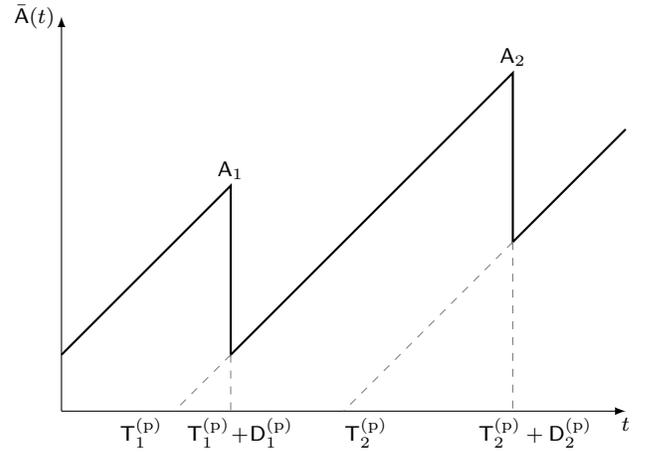
\begin{figure}[t]
  \centering
\begin{tikzpicture}[font=\footnotesize,scale=0.75]
	\draw[-latex] (0,0) -- (0,7);
	\draw[-latex] (0,0) -- (10,0);
	\draw (10,0) node[below]{$t$};
	\draw (0,7) node[left]{$\ageofinfo(t)$};
	\draw[thick] (0,1) -- (3,4) -- (3, 1) -- (8,6) -- (8,3) -- (10,5);
	\draw[gray,dashed] (3,1) -- (2,0);
	\draw[gray,dashed] (3,1) -- (3,0);
	\draw[gray,dashed] (8,3) -- (5,0);
	\draw[gray,dashed] (8,3) -- (8,0);
	\draw (1.5,0) node[below]{$\arrivaltimeofpacket{1}\hspace{1ex}$};
	\draw (3,0) node[below]{$\hspace{2ex} \arrivaltimeofpacket{1}\!+\!\delayofpacket{1}$};
	\draw (5.4,0) node[below]{$\arrivaltimeofpacket{2}$};
	\draw (8.4,0) node[below]{$\arrivaltimeofpacket{2}+\delayofpacket{2}$};
	\draw (3,4) node[above]{$\peakageofinfocountat{1}$};
	\draw (8,6) node[above]{$\peakageofinfocountat{2}$};
\end{tikzpicture}
\caption{Peak age of information for the frame-synchronous model: $\arrivaltimeofpacket{m}$ is the frame index corresponding to the arrival of the $m$th transmitted packet; $\arrivaltimeofpacket{m}+\delayofpacket{m}$ is the frame index at which the $m$th transmitted packet leaves the queue; the peak age $\peakageofinfocountat{m}$ is the age of information just before the $m$th transmitted packet departs.}\label{fig:peak_age}
\end{figure}

Let now $\arrivaltimeofpacket{m}+\delayofpacket{m}$ be the time frame at which the $m$th transmitted packet leaves the queue.
It follows that the index of the most recent packet received at the destination at time frame $t$ is~$\widehat{m}(t)=\max\biggl\{ m: \arrivaltimeofpacket{m}+\delayofpacket{m}\leq t \biggr\}$
and the corresponding time stamp is $\arrivaltimeofpacket{\widehat{m}(t)}$.
The age of information is the discrete-time random process~$\ageofinfo(t)= t- \arrivaltimeofpacket{\widehat{m}(t)}, \, t=1,2,\dots$;
the peak-age of information $\peakageofinfocountat{m}$ is the  value of the age of information just before the update contained in the $m$th transmitted packet is received (see Fig.~\ref{fig:peak_age}).
As shown in the figure, the peak age $\peakageofinfocountat{m}$ is the sum of the delay $\delayofpacket{m}$ of the $m$th transmitted packet and of the difference $\arrivaltimeofpacket{m}-\arrivaltimeofpacket{m-1}$ between the frame indices corresponding to the arrival of the $(m-1)$th and the $m$th transmitted packets.
%
%
Similar to~\eqref{expr:delayviolation_criteria}, we define the steady-state peak-age violation probability as
\begin{equation}
    \Bprobofpeakageviolationactual{\peakagethreshold} = \limsup_{m\tendsto\infty} \left[1- \probof{\peakageofinfocountat{m}<\peakagethreshold/\framesize\text{ and } \messageofpacket{m}\neq\decodedmessageofpacket{m}}\right].
\end{equation}
Then, we upper-bound it as
\begin{IEEEeqnarray}{rCCCl}
\Bprobofpeakageviolationactual{\peakagethreshold} &\leq & \probof{\peakageofinfo\geq\peakagethreshold/\framesize}+\packeterrorprob&\triangleq& \Bprobofpeakageviolation{\peakagethreshold}. \label{expr:peakageviolation_criteria}
\end{IEEEeqnarray}
Here, $\peakageofinfo$ is steady-state peak age of information, defined as $\peakageofinfo =  \delay + \arrivaltimeofbulk{2}-\arrivaltimeofbulk{1}.$
Note that we dropped the superscript $\mathrm{(p)}$ to keep the notation compact.
Our analysis of the delay violation  probability in Section~\ref{sec:steadystate_delay} assumes that the queue operates according to a FCFS discipline.
When packets contain status updates and one is interested in the freshness of the information available at the receiver, it is natural to consider other queue management policies that are better suited to minimize the peak-age violation probability.

In what follows, we analyze the steady-state peak-age violation probability for the three packet management policies proposed in~\cite{costa2016age} in the context of peak-age analyses under continuous-time exponential arrival and service processes: \newtext{ (i) FCFS with system capacity 1 (\DWTname); (ii) FCFS with system capacity 2 (\KTNname); and (iii) last come first serve with preemption in queue (\KTLname)}.
The \newtext{\DWTname} policy discards every newly arrived packet during the transmission of a given packet.
Under such a policy, the \newtext{system} can contain at most one packet \newtext{at any given time}.
%
In the \newtext{\KTNname} and the \newtext{\KTLname} policies, the queue is allowed to hold an additional packet other than the one currently served.
In \newtext{\KTNname}, this additional packet is the first packet arrived in a new frame during the service time of the current packet.
In \newtext{\KTLname}, it is the last packet arrived in a new frame.
These two policies can be modeled by a system with maximum capacity of $2$ and nonpreemptive/preemptive buffering, respectively~\cite{rubin1988performance}.
\newtext{We also consider a new policy, which we denote as LCFS with preemption in service (LCSF-S).
According to this policy, whenever a new packet arrives, the transmission of the current packet is interrupted and the transmission of the new packet is started in the next frame.}

\newtext{The policies \DWTname and \KTNname involve no preemption; hence, they may be more suitable for energy-constrained applications.
\KTLname achieves lower peak age violation probability than \KTNname because it allows for preemption.
However, its queue management is more computationally intensive because a full queue needs to be updated whenever a new packet arrives.
Finally, LCSF-S is the most complex queue-management policy, but, as we shall see, it also offers the best performance and appears to be best suited to URLLC.}

We shall next derive the PGF of the steady-state peak age for \newtext{the four} packet management policies.
The corresponding peak-age violation probability can be obtained by following the steps detailed in Section~\ref{sec:steadystate_delay}.
We start with the \newtext{\DWTname} policy.
\begin{thm}\label{thm:pgf_steadystate_peakage_dwt}
For a given packet interarrival rate $\packetarrivalprob$ and for every~$\parantheses{\framesize,\infobits,\probservicestoppingtime,\packeterrorprob}$--VLSF code satisfying the stability condition $\packetarrivalprob\framesize\Bexpectation{\servicestoppingtime}<1$, the PGF of the peak age of information $\peakageofinfo$ at steady state with \newtext{\DWTname} for the frame-synchronous model is
\begin{IEEEeqnarray}{rCl}
 \BPGFof{\peakageofinfo}{s} &=& {\BPGFof{\servicestoppingtime}{s}^2s\parantheses{1-\parantheses{1-\packetarrivalprob}^\blocklength}}/\parantheses{\parantheses{1-s\parantheses{1-\packetarrivalprob}^\blocklength}}.\label{expr:pgf_steadystate_peakage_dwt} \IEEEeqnarraynumspace
\end{IEEEeqnarray}
\end{thm}
\begin{IEEEproof}
    Since the packets that arrive during the service time of a given packet are discarded, the interarrival time $\arrivaltimeofbulk{2}-\arrivaltimeofbulk{1}$ is no longer geometrically distributed.
It is then convenient to express the peak age as $\peakageofinfo=\servicetimeofpacket{1}+\servicetimeofpacket{2}+\arrivaltimeofbulk{}$ where $\servicetimeofpacket{1}$ and $\servicetimeofpacket{2}$ are the service time of two packets and $\arrivaltimeofbulk{}$ is the amount of time (measured in number of frames) elapsed from the completion of the service of the first packet until the next packet arrives.
These three random variables are independent.
Furthermore, $\arrivaltimeofbulk{}\distas\Bgeometricdist{1-\parantheses{1-\packetarrivalprob}^\blocklength}$ in the frame-synchronous case.
We obtain~\eqref{expr:pgf_steadystate_peakage_dwt} by using that the random variables $\servicetimeofpacket{1}$ and $\servicetimeofpacket{2}$ have probability distribution $\probservicestoppingtime$.
\end{IEEEproof}
%

\begin{figure}
\iftwocol
\centering\includegraphics[scale=0.99]{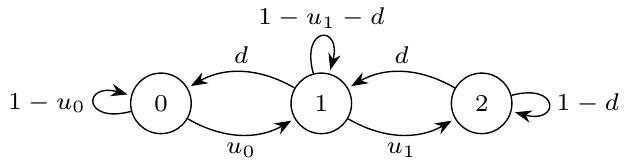}
\else
\centering\includegraphics[scale=1.2]{MM12MarkovChain}
\fi
\caption{
State transition diagram of the size of the queue for the \newtext{\KTNname} and \newtext{\KTLname} packet management policies.
}\label{fig:mm_1_2_queue_size_markov_chain}
\end{figure}

An analysis of the remaining \newtext{three} polices under general VLSF codes appears unfeasible.
Hence, we shall focus in the reminder of this section on the simpler case of ARQ with perfect error detection.
Under these assumptions, the service time follows a geometric distribution.
The \newtext{memorylessness} of the geometric distribution allows us to adapt to our scenario the derivation presented in~\cite[Secs. IV-B and IV-C]{costa2016age} for the continuous-time case under an abstract model for the service time.
This is done \newtext{for \KTNname and \KTLname} in the next theorem.\footnote{ \newtext{
The proof, which follows from an adaptation to   the discrete-time case of the analysis in~\cite[Secs. IV-B and IV-C]{costa2016age}  is omitted due to space constraint.}}
\begin{thm}\label{thm:peakage_KTN_KTL_ARQ_sync}
Consider the ARQ frame-synchronous model.
The PGF of the peak age of information for  \newtext{\KTNname} and \newtext{\KTLname} is
\iftwocol
\begin{IEEEeqnarray}{rCl}
\BPGFof{\peakageofinfo}{s} &=& \BPGFof{\waitingtime}{s}\BPGFof{\servicetimeofpacket{} }{s}\nonumber\\
&&.\parantheses{ \frac{p_0}{p_0+p_1}\BPGFof{\servicetimeofpacket{}^{(0)} }{s}\BPGFof{\arrivaltime}{s} + \frac{p_1}{p_0+p_1}\BPGFof{\servicetimeofpacket{}^{(1)} }{s}}\IEEEeqnarraynumspace\label{expr:peakage_KTN_KTL_ARQ_sync}
\end{IEEEeqnarray}
\else
\begin{IEEEeqnarray}{rCl}
\BPGFof{\peakageofinfo}{s} &=& \BPGFof{\waitingtime}{s}\BPGFof{\servicetimeofpacket{} }{s}\parantheses{ \frac{p_0}{p_0+p_1}\BPGFof{\servicetimeofpacket{}^{(0)} }{s}\BPGFof{\arrivaltime}{s} + \frac{p_1}{p_0+p_1}\BPGFof{\servicetimeofpacket{}^{(1)} }{s}}.\IEEEeqnarraynumspace\label{expr:peakage_KTN_KTL_ARQ_sync}
\end{IEEEeqnarray}
\fi
Here, $\arrivaltime\distas\Bgeometricdist{1-\parantheses{1-\packetarrivalprob}^\blocklength}$
with PGF
\begin{IEEEeqnarray}{rCl}
\BPGFof{\arrivaltime}{s} &=& {s\parantheses{1-\parantheses{1-\packetarrivalprob}^\framesize}}\parantheses{1-\parantheses{1-\packetarrivalprob}^\framesize s}\label{expr:pgf_arrival_bulk}
\end{IEEEeqnarray}
is the amount of time elapsed from the completion of the service of a packet  until the next packet arrives; $\waitingtime$ is the waiting time of a packet, whose PGF is given by
\begin{IEEEeqnarray}{rCl}
\BPGFof{\waitingtime}{s} &=&
\begin{cases}
\displaystyle
\frac{p_0}{p_0+p_1} + \frac{p_1\BPGFof{\servicetimeofpacket{}}{s}}{\parantheses{p_0+p_1}s}, &\quad \text{for \newtext{\KTNname}} \label{expr:waitingtime_KTN_ARQ_sync}\\[6mm]
 \displaystyle p_0 + \parantheses{1-p_0}\BPGFof{\servicetimeofpacket{}^{(0)} }{s}/{s} ,& \quad \text{for \newtext{\KTLname}}.
\end{cases}
\end{IEEEeqnarray}
%
In~\eqref{expr:peakage_KTN_KTL_ARQ_sync} and~\eqref{expr:waitingtime_KTN_ARQ_sync},
\begin{IEEEeqnarray}{rCl}
p_0 &=& {d^2}/\parantheses{d^2+u_0d+u_0u_1}\label{expr:ktn_ktl_steady_state_queue_empty}
\end{IEEEeqnarray}
where
\begin{IEEEeqnarray}{rCl}
u_0 &=& 1-\parantheses{1-\packetarrivalprob}^\framesize \label{expr:ktn_ktl_queue_jump_up_0}\\
u_1 &=& \arqblockerrorprob\parantheses{1-\parantheses{1-\packetarrivalprob}^\framesize}\label{expr:ktn_ktl_queue_jump_up_1}\\
d &=& \parantheses{1-\arqblockerrorprob}\parantheses{1-\packetarrivalprob}^\framesize\label{expr:ktn_ktl_queue_jump_down}.
\end{IEEEeqnarray}
Furthermore,
\begin{IEEEeqnarray}{rCl}
p_1 &=& {u_0p_0}/{d}\label{expr:ktn_ktl_steady_state_queue_one}\\
\BPGFof{\servicetimeofpacket{} }{s} &=& s\parantheses{1-\arqblockerrorprob}/\parantheses{1-\arqblockerrorprob s}\label{expr:ktn_ktl_steady_state_service}\\
\BPGFof{\servicetimeofpacket{}^{(0)} }{s} &=& s{\parantheses{1-\arqblockerrorprob\parantheses{1-\packetarrivalprob}^{\framesize}}}/\parantheses{1-\arqblockerrorprob\parantheses{1-\packetarrivalprob}^{\framesize} s}\label{expr:ktn_ktl_steady_state_service_no_arrivals}\\
\BPGFof{\servicetimeofpacket{}^{(1)} }{s} &=& \parantheses{\parantheses{p_0+p_1}\BPGFof{\servicetimeofpacket{} }{s}-p_0\BPGFof{\servicetimeofpacket{}^{(0)} }{s}}/{p_1}. \label{expr:ktn_ktl_steady_state_service_one_arrivals} \IEEEeqnarraynumspace
\end{IEEEeqnarray}
\end{thm}
Some remarks are in order.
As shown in~\cite{costa2016age}, in both \newtext{\KTNname} and \newtext{\KTLname} the queue size can be described by the Markov chain with three states depicted in Fig.~\ref{fig:mm_1_2_queue_size_markov_chain}.
The state transition probability $u_0$ in~\eqref{expr:ktn_ktl_queue_jump_up_0} is the probability that a new packet arrives in a given frame, $u_1$ in~\eqref{expr:ktn_ktl_queue_jump_up_1} is the probability that a new packet arrives and  that a NACK is fed back in a given frame, and $d$ in~\eqref{expr:ktn_ktl_queue_jump_down} is the probability that no new packet arrives and an ACK is fed back in a given frame.
The parameters $p_0$ and $p_1$ in~\eqref{expr:ktn_ktl_steady_state_queue_empty} and~\eqref{expr:ktn_ktl_steady_state_queue_one} are the probabilities that the queue has size zero and one in steady state, respectively.
Finally, in~\eqref{expr:ktn_ktl_steady_state_service}--\eqref{expr:ktn_ktl_steady_state_service_one_arrivals}, $\servicetimeofpacket{}\distas \probservicestoppingtime$ is the service time and $\servicetimeofpacket{}^{(0)}$ and $\servicetimeofpacket{}^{(1)}$ are the conditional service time given that no new packet and one new packet has entered the queue, respectively.
%

\newtext{Finally, we characterize PGF of the peak age of information for \ATLname in the following theorem.
\begin{thm}\label{thm:peakage_ATL_ARQ_sync}
Consider the ARQ frame-synchronous model.
The PGF of the peak age of information for   \ATLname is
\iftwocol
\begin{IEEEeqnarray}{rCl}
\BPGFof{\peakageofinfo}{s} &=& \BPGFof{\arrivaltimeofbulk{}^{(0)}}{s} \frac{\packetdeliveredprob\BPGFof{\servicetimeofpacket{}^{(0)}}{s}}{1-\parantheses{1-\packetdeliveredprob}\BPGFof{\servicetimeofpacket{}^{(0)}}{s}}.\IEEEeqnarraynumspace\label{expr:peakage_ATL_ARQ_sync}
\end{IEEEeqnarray}
\else
\begin{IEEEeqnarray}{rCl}
\BPGFof{\peakageofinfo}{s} &=& \BPGFof{\arrivaltimeofbulk{}^{(0)}}{s} \frac{\packetdeliveredprob\BPGFof{\servicetimeofpacket{}^{(0)}}{s}}{1-\parantheses{1-\packetdeliveredprob}\BPGFof{\servicetimeofpacket{}^{(0)}}{s}}.\IEEEeqnarraynumspace\label{expr:peakage_ATL_ARQ_sync}
\end{IEEEeqnarray}
\fi
where~$\BPGFof{\servicetimeofpacket{}^{(0)} }{s}$ is defined in~\eqref{expr:ktn_ktl_steady_state_service_no_arrivals},
\begin{IEEEeqnarray}{rCl}
\packetdeliveredprob &=& \parantheses{1-\arqblockerrorprob}/\parantheses{1-\arqblockerrorprob\parantheses{1-\packetarrivalprob}^{\framesize}}\label{expr:atl_pacekt_delivery_prob}\\
\BPGFof{\arrivaltimeofbulk{}^{(0)} }{s} &=& \parantheses{\BPGFof{\arrivaltimeofbulk{} }{s}-\parantheses{1-\packetdeliveredprob}\BPGFof{\servicetimeofpacket{}^{(0)} }{s}}/{\packetdeliveredprob}. \label{expr:atl_steady_state_service_one_arrivals}
\end{IEEEeqnarray}
In~\eqref{expr:atl_steady_state_service_one_arrivals}, the PGF~$\BPGFof{\arrivaltimeofbulk{} }{s}$ is given in~\eqref{expr:pgf_arrival_bulk}.
\end{thm}
\begin{IEEEproof}
  See Appendix~\ref{proof:peakage_ATL_ARQ_sync}.
\end{IEEEproof}
Some comments on Theorem~\ref{thm:peakage_ATL_ARQ_sync} are in order.
The constant $\packetdeliveredprob$ in~\eqref{expr:atl_pacekt_delivery_prob} is the probability that no preemption in service occurs.
The random variable~$\servicetimeofpacket{}^{(0)}$ is distributed as the conditional  service time given that no preemption occurs.
The random variable~$\arrivaltimeofbulk{}$  denotes the number of frames elapsed between two consecutive packet arrivals; finally, the distribution of the random variable~$\arrivaltimeofbulk{}^{(0)}$ equals the conditional distribution of $\arrivaltimeofbulk{}$ given that no preemption occurs.
}

\section{Numerical Results}\label{sec:numerical_results}
\newtext{Throughout this section, we consider a binary-input AWGN channel with input $x\in\{ -\sqrt{\snr},\sqrt{\snr}\}$ for $\snr>0$, output $y\in\reals$ and channel law
$P_{\chout\given\chinp}\parantheses{\Rchout\given\Rchinp} = \frac{1}{\sqrt{2\pi}}\upnegexp{\frac{(\Rchout-\Rchinp)^2}{2}}$.
Note that, since the variance of the Gaussian additive noise is one, the parameter~$\snr$ defines the signal-to-noise ratio (SNR).
This choice allows us to focus on the impact of key parameters such as number of information bits, frame size, etc., without the need to make additional choices such as pilot design for fading channels or antenna processing for multiantenna systems. Nevertheless, it is important to highlight that our analysis holds for more general channel models that include fading, multiple antennas, pilot-assisted transmission, and mismatch decoding as well as interference.
Such extensions, which are relevant for URLLC, are discussed in Section~\ref{sec:fading_interference}.
}

We provide next  numerical investigations aimed at determining the dependence of the delay violation and the peak-age violation probability on system parameters such as SNR, frame size, undetected error probability as well as packet-management policy for the case of age of information.
We will also illustrate the tightness of the saddlepoint approximation~\eqref{expr:ccdf_steadystate_delay_saddlepoint} and discuss, for  ARQ with perfect error detection, the accuracy of the stochastic-network calculus bound given in Theorem~\ref{thm:ccdf_steadystate_delay_netcalc}.
Throughout, we will assume that the information packets carry $\infobits=30$ bits \newtext{unless specified otherwise}.
For space constraint, we shall focus uniquely on the frame-synchronous setup.
\paragraph*{Simple ARQ with Perfect Error Detection}
\begin{figure}
\centering
\includegraphics[width=0.85\columnwidth]{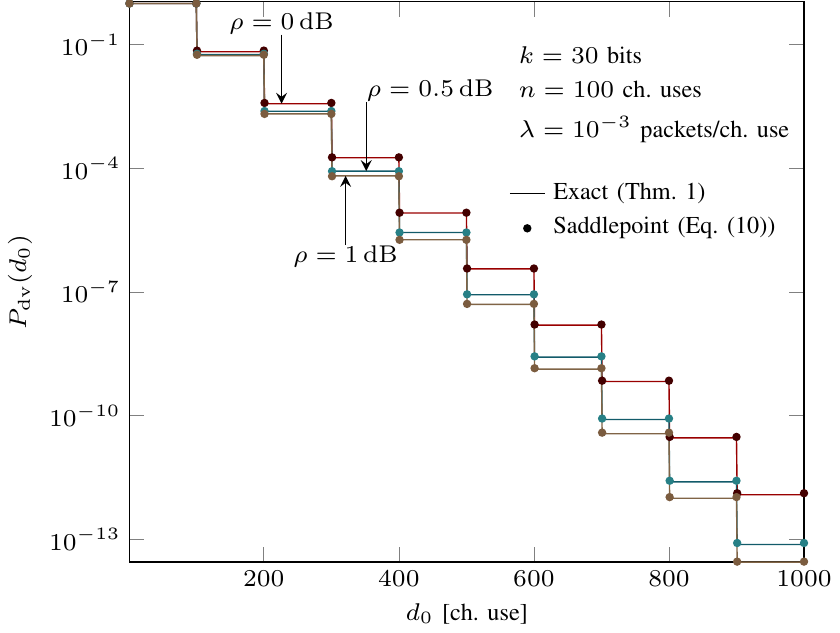}
\caption{Delay violation probability for different values of SNR~$\snr$ for the case of ARQ with perfect error detection.}\label{fig:delayviolation_sync_arq_vary_snr}
\end{figure} 
In Fig.~\ref{fig:delayviolation_sync_arq_vary_snr}, we study the dependence of the delay violation probability $\Bprobofdelayviolationactual{\delaythreshold}$ on the SNR $\snr$ for the case of ARQ with perfect error detection.
We assume a frame size~$\framesize=100$ channel uses and an arrival rate $\averagearrivalrate=10^{-3}$ packets/channel use.
The delay violation probability is computed using both a recursion-based $z$ transform inversion, which is used to evaluate~\eqref{expr:ccdf_steadystate_delay} and the reduced-complexity saddlepoint approximation~\eqref{expr:ccdf_steadystate_delay_saddlepoint}.
As one can see from the figure, the saddlepoint approximation is extremely accurate.
This turns out to be true also in the more general VLSF setup.
When the SNR increases, the channel becomes more reliable and,
hence, both the service time and the overall packet delay decrease.

\begin{figure}
\centering
\includegraphics[width=0.85\columnwidth]{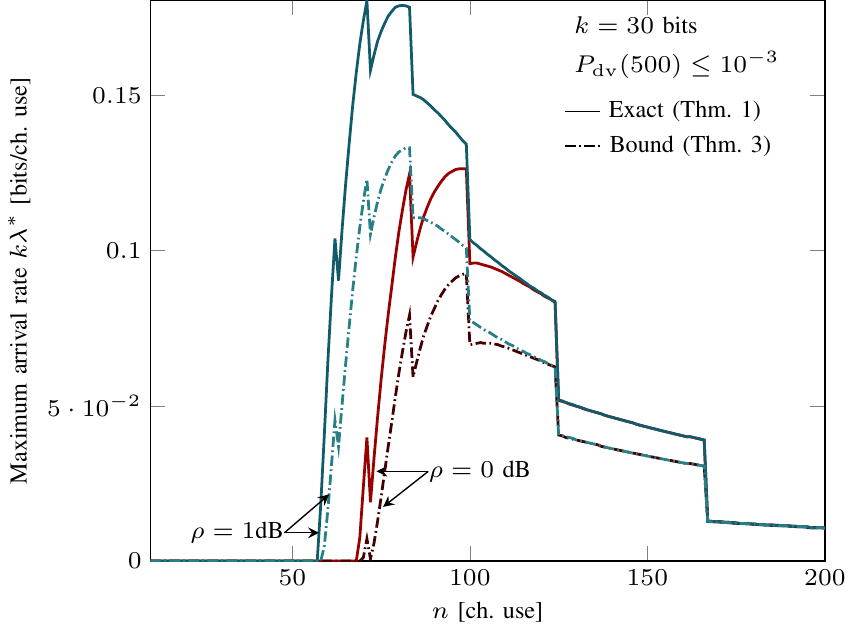}
\caption{Maximum throughput as a function of the frame size for ARQ with perfect error detection.}\label{fig:maxarrvial_constraint_delayviolation_sync_arq_vary_snr}
\end{figure} 

Next, we study the maximum throughput achievable under a constraint on the delay violation probability.
The throughput is defined as the product $\infobits\averagearrivalrate^*$ between the number of information bits per packet $k$ and the maximum packet arrival rate $\averagearrivalrate^*$ .
In Fig.~\ref{fig:maxarrvial_constraint_delayviolation_sync_arq_vary_snr}, we plot the throughput $\infobits\averagearrivalrate^*$  as a function of the frame size $\framesize$ for different values of $\snr$.
In the figure, we set $\delaythreshold=500$ channel uses and a target delay violation probability of $10^{-3}$.
As a reference, we also plot throughput estimates obtained by using the upper bound on the delay violation probability~\eqref{expr:ccdf_steadystate_delay_netcalc}, which relies on stochastic network calculus.
From the figure we see that, for the case $\snr=0\dB$, the throughput estimate based on~\eqref{expr:ccdf_steadystate_delay_netcalc} is about $35\%$ lower than what predicted by our exact analysis.
The figure illustrates that there exists a throughput maximizing frame size.
In fact, on the one hand, when the frame size is small, the packet error probability  is large and so is the number of retransmissions, yielding a large probability that the delay exceeds the threshold for a given arrival rate.
On the other hand, when the frame size is large, the packet error probability is small, but even a small number of retransmissions is sufficient to generate a large delay for a given arrival rate.
Note that the discontinuities in the plot, which occur at submultiples of $\delaythreshold$, are caused by the change in the number of available retransmission rounds.
Finally, although the bound~\eqref{expr:ccdf_steadystate_delay_netcalc} provides a loose throughput estimate, it seems to predict accurately the value of the throughput-maximizing frame size.

\paragraph*{General VLSF Codes}
\begin{figure}
\centering
\includegraphics[width=0.85\columnwidth]{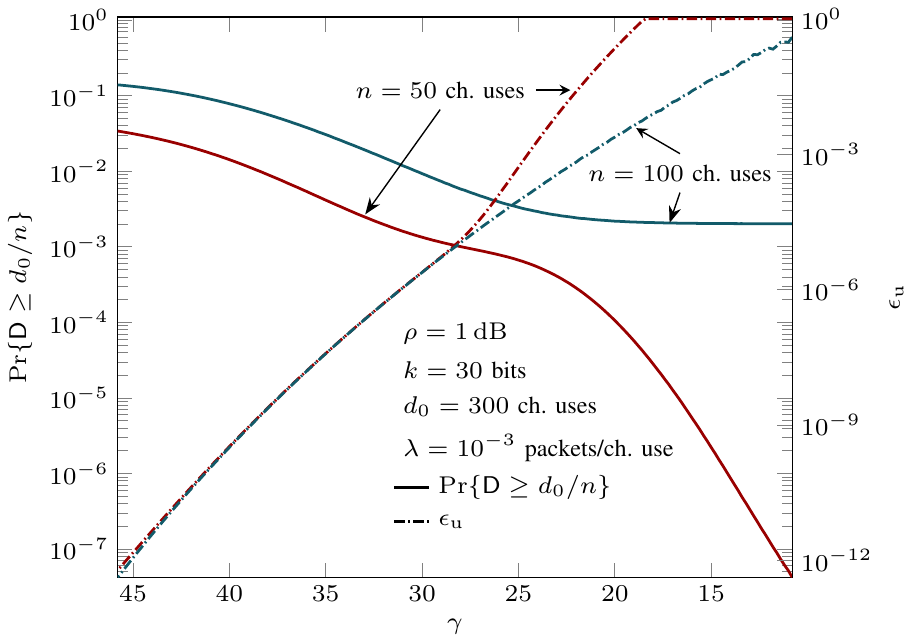}
\caption{Steady state delay CCDF and undetected error probability for different values of threshold~$\infodensitythreshold$ for general VLSF setup.}\label{fig:delayviolation_and_errorprob_vs_gamma}
\end{figure} 
We now consider VLSF codes combined with the threshold decoder used in the proof of Theorem~\ref{thm:vlsf_achievability}.
Fig.~\ref{fig:delayviolation_and_errorprob_vs_gamma} plots the CCDF of the steady-state delay and the undetected packet error probability $\packeterrorprob$ as a function of the decoding threshold $\infodensitythreshold$.
We recall that the upper bound $\Bprobofdelayviolation{\delaythreshold}$ on the delay violation probability consists of the sum of these two quantities---see~\eqref{expr:delayviolation_criteria_bound}.
When the threshold $\infodensitythreshold$ is low, the probability that the delay exceeds the latency constraint is small, because the decoding process terminates early, but the undetected error probability is high.
In fact, it is likely that a codeword different from the transmitted one has an accumulated information density that exceeds the threshold.
In contrast, as the threshold $\infodensitythreshold$ increases, the undetected error probability decreases but the probability that the delay is above threshold increases.



\begin{figure}
\centering
\includegraphics[width=0.85\columnwidth]{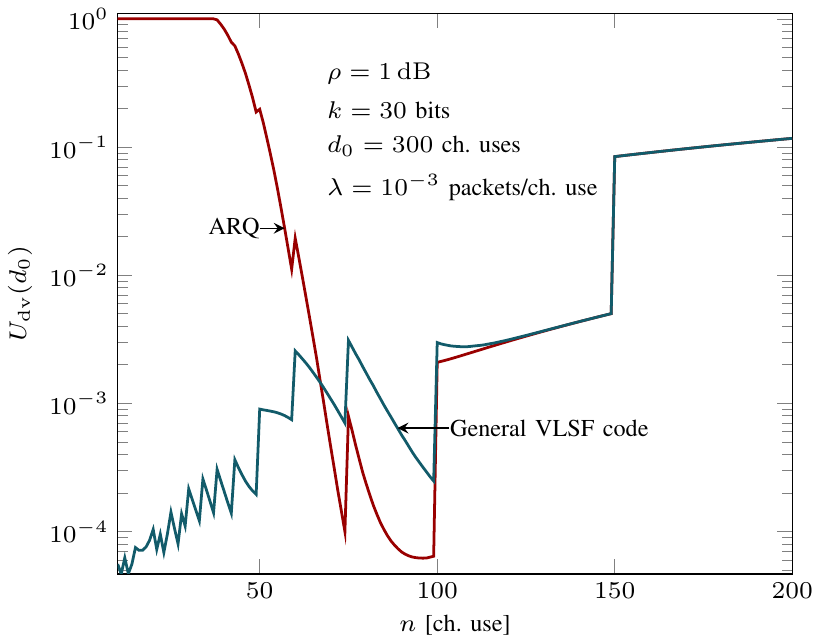}
\caption{Delay violation probability versus frame size for both  ARQ with perfect error detection and general VLSF setup. }\label{fig:delayviolation_vs_blocklength_arq_vs_harq}
\end{figure} 

In Fig.~\ref{fig:delayviolation_vs_blocklength_arq_vs_harq}, we
plot the delay violation probability versus the frame size for both ARQ with perfect error detection and the general VLSF setup.
Note that the comparison is not fair because of the perfect-error detection assumption in the ARQ setup.
Yet, one sees that for small frame sizes,  VLSF codes  yield  lower delay violation probability.
This is because choosing the frame size too small or too big penalizes performance of ARQ, whereas decreasing the frame size enhances the performance of VLSF since it allows the decoder to sample the accumulated information density more frequently.


Finally, we elaborate on the difference between optimizing a system for a target average delay and optimizing it for a target delay violation probability.
To this end, we consider  ARQ with perfect error detection  and focus on the parameters in Fig.~\ref{fig:delayviolation_vs_blocklength_arq_vs_harq}.
For these parameters, the frame size values of~$\framesize=60$ channel uses  and~$\framesize=91$ channel uses result in very similar average delays, namely about $95.8$ and~$96$ channel uses, respectively.
However, they yield significantly different delay violation probabilities, namely about $1.38\times 10^{-2}$ and~$6.44\times 10^{-5}$, respectively.
This highlights the importance of performing delay violation probability analyses in latency-critical wireless systems.



\paragraph*{Peak-Age Violation Probability}
\begin{figure}
\centering
\includegraphics[width=0.85\columnwidth]{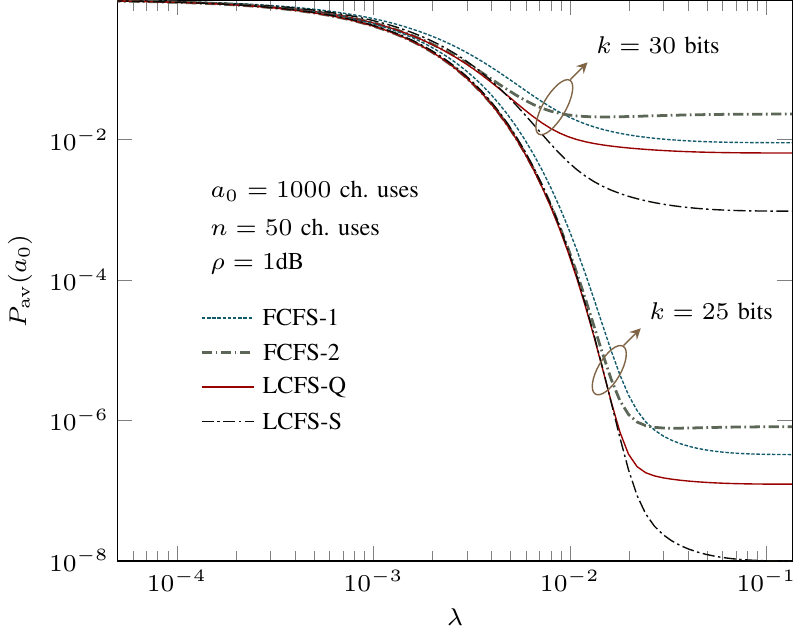}
\caption{Peak-age violation probability as a function of the packet arrival rate for different packet-management policies.}\label{fig:ageviolation_vs_arrivalrate_policies}
\end{figure} 
In Fig.~\ref{fig:ageviolation_vs_arrivalrate_policies}, we illustrate the peak-age violation probability for the case of frame-synchronous system and ARQ with perfect error detection, as a function of the packet arrival rate $\averagearrivalrate$, for the packet-management policies analyzed in Section~\ref{sec:age}, namely \newtext{\DWTname}, \newtext{\KTNname}, \newtext{\KTLname}, and \newtext{\ATLname}.
\newtext{We consider both~$\infobits=25$ bits and~$\infobits=30$ bits; as expected, the peak age violation probability increases with number of information bits.}
We also observe that \newtext{\ATLname} is uniformly better than \newtext{\DWTname}, \newtext{\KTNname}, and \newtext{\KTLname}.
Furthermore, \newtext{\DWTname} outperforms \newtext{\KTNname} at high arrival rates, whereas the situation is reversed at lower arrival rates.
Indeed, for high arrival rates,~$\Bprobofpeakageviolationactual{\peakagethreshold}$ converges to%
\begin{IEEEeqnarray}{rCl}
  \IEEEeqnarraymulticol{3}{l}{ \lim_{\averagearrivalrate\tendsto 1}\Bprobofpeakageviolationactual{\peakagethreshold}  } \nonumber \\
&=& \begin{cases}
 \probof{\servicetimeofpacket{1}+\servicetimeofpacket{2}+1\geq \peakagethreshold/\framesize} , &\quad \text{for \newtext{\DWTname}}\\
\probof{\servicetimeofpacket{1}+\servicetimeofpacket{2}+\servicetimeofpacket{3}-1\geq \peakagethreshold/\framesize}, &\quad \text{for \newtext{\KTNname}} \\
 \probof{\servicetimeofpacket{1}+\servicetimeofpacket{2}\geq \peakagethreshold/\framesize} , &\quad \text{for \newtext{\KTLname}}\\
 \newtext{\probof{\servicetimeofpacket{1}+1\geq \peakagethreshold/\framesize}} , &\quad \text{for \newtext{\ATLname}}
\end{cases}\nonumber%
\end{IEEEeqnarray}%
where $\servicetimeofpacket{1},\servicetimeofpacket{2},$ and $\servicetimeofpacket{3}$ are \iid and~$\servicetimeofpacket{1}\sim\Bgeometricdist{1-\arqblockerrorprob}$.
For the case \newtext{$\infobits=25$ bits, this limit is~$3.2\times 10^{-7}$ for \newtext{\DWTname}, $8\times 10^{-7}$ for \newtext{\KTNname}, $1.2\times 10^{-7}$ for \newtext{\KTLname}, and \newtext{$9.7\times 10^{-9}$ for \newtext{\ATLname}}}.
These results support the findings in~\cite{costa2016age} \newtext{and extend them to a more realistic physical-layer model as well as to a policy, LCFS-S, that is more suitable to URLLC}.


\section{Extensions of the Framework}
\subsection{Fading and Interference}\label{sec:fading_interference}
\newtext{While the results in the previous section focus on the standard bi-AWGN channel, the analysis can be easily extended to encompass more realistic system models for URLLC transmission, which include fading, multiple antennas, pilot-assisted transmission, and mismatch nearest-neighbor decoding.
For example, a key enabler of URLLC is frequency diversity, which can be captured in our framework using a block-fading model in which each transmission frame spans a number of resource blocks in the time-frequency plane that are separated in frequency by more than the channel coherence bandwidth.
The resulting channel is not memoryless, but rather block memoryless.
Our analysis can be generalized to this setup by replacing the finite-blocklength bounds in Theorem~\ref{thm:rcus} and~\ref{thm:vlsf_achievability} with their block-memoryless counterparts (see e.g.,~\cite{ostman2018lowglobecompublished,ostman2018short}).
Pilot-assisted transmission for channel estimation and mismatch nearest-neighbor decoding at the receiver can be also accounted for in the analysis~\cite{ostman2018lowglobecompublished,ostman2018short}.
Note that the inclusion of mismatch nearest-neighbor decoding allows one to characterize the impact on performance of an unknown interference signal, when such a signal is treated by the decoder as an additive noise.
}

\subsection{Beyond the Point-to-Point Setup}\label{sec:beyond_p2p}
\newtext{Our analysis pertains a point-to-point link serving a single flow of data.
The following are natural generalizations: (i) a point-to-point link serving multiple independent flows of information, possibly with different reliability, freshness, and latency requirements or priorities as in \cite{TelatarNajm,popovski20185Gslicing}; (ii) a broadcast scenario with common information to be sent to all receivers; (iii) a broadcast scenario where independent information flows are to be sent to different receivers; (iv) the multiple-access case, where multiple sensors send information to a common destination; (v) the multi-hop setting where the objective is to maintain low end-to-end latency and high end-to-end data freshness.}

\newtext{The first three scenarios can be addressed by generalizing the framework presented in this paper as discussed below.}
\newtext{For scenario (i), the steady-state average age of information under LCFS-S and LCFS-Q was derived in~\cite{YatesKaul2017}.
An analysis of the delay and the peak age violation probabilities could be tackled using the tools we presented in this paper.
To analyze scenario (ii), one can leverage the nonasymptotic bounds presented in~\cite{trillingsgaard2018common} on the performance of VLSF codes on the broadcast channel with common message.
A special case of scenario (iii), in which at each time the transmitter chooses a single flow to serve, was recently analyzed in~\cite{Kadota2018}.
For the case of deterministic packet arrivals, one can leverage the indexability of the problem.
For more general arrival distributions, the multi-server results in \cite{BedewyMultihop2017} on the optimality of the last-generated first-serve packet management policy can be leveraged.
All these analyses rely on a simplified physical-layer model and can be combined with the framework introduced in this paper to obtain more accurate performance predictions, and derive guidelines for a joint design of physical and higher layers.}

\section{Conclusion}\label{sec:conclusion}
The optimal design of URLLC must rely on nonasymptotic joint coding-queuing analyses that target tail probabilities rather than averages.
In this paper, we illustrated how to perform such an analysis for the case of \iid Bernoulli packet arrivals, single-server queue, and point-to-point communications over a bi-AWGN channel via VLSF codes.
We presented methods for evaluating the delay violation and the peak-age violation probability and analyze the dependence of these two quantities on system parameters such as the frame length, the undetected-error probability and the packet-management policy.
A novel aspect of our approach is that it accounts also for undetected errors.
Such errors may be extremely harmful in mission-critical communication systems and specific designs aimed at minimizing them will be the subject of future works.

Our analysis illustrates that the delay violation probability is extremely sensitive to the value of the system parameters.
This can be seen in Figs.~\ref{fig:maxarrvial_constraint_delayviolation_sync_arq_vary_snr} and~\ref{fig:delayviolation_vs_blocklength_arq_vs_harq}, which show the dependence of the maximum throughput and of the delay violation probability on the frame size for the case of simple ARQ with perfect error detection, and in Fig.~\ref{fig:delayviolation_and_errorprob_vs_gamma}, which depicts the dependence of the delay violation probability on the undetected error probability.
We also showed that analyses based on average latency or on a stochastic-network-calculus approach do not yield an optimal design and, for the case of stochastic network calculus, suffer from lack of generality.


\begin{appendices}

  \section{Proof of Theorem~\ref{thm:delay_sync_steadystate}}\label{proof:delay_sync_steadystate}
  Let us denote by $\queuesizepktleave_m$ the number of bulks remaining in the buffer just after the $m$th bulk leaves the buffer, i.e.,~$\queuesizepktleave_m = \queuesizeattime{\arrivaltimeofbulk{m}+\delayofbulk{m}}$.
 We observe that $\curlybrac{\queuesizepktleave_m}_{m=1}^{\infty}$,  is a Markov chain governed by~$\queuesizepktleave_{m+1} =
      \max\{\queuesizepktleave_m - 1,0\} + \arrivalsduringserving_m$
  where $  \arrivalsduringserving_m =\sum_{t=1}^{\servicetimeofbulk{m+1}} \indicator{\bulkarrivalat{\Bmax{\arrivaltimeofbulk{m}+\delayofbulk{m},\arrivaltimeofbulk{m+1}}+t}>0}$
  %
  is the number of bulks of packets arriving during the service time of the $m$th bulk.
  Note that $\arrivalsduringserving_m$ depends only on the length of the service time $\servicetimeofbulk{m+1}$ and is independent of $\queuesizepktleave_m$.
  Hence,~$\arrivalsduringserving_m \distas \sum_{t=1}^{\servicetimeofbulk{1}} \indicator{\bulkarrivalat{t}>0}$.
  The PGF of the steady-state buffer-size $\queuesizepktleave$ is~$\BPGFof{\queuesizepktleave}{s} = \parantheses{1-\Bexpectation{\arrivalsduringserving}}(s-1)\BPGFof{\arrivalsduringserving}{s}/\parantheses{s-G_{\arrivalsduringserving}(s)},\ \Bexpectation{\arrivalsduringserving}<1$~\cite[Eq. (11.3.11)]{grimmett2001probability}.
  Since $\queuesizepktleave \sim \sum_{t=1}^{\delay}\indicator{\bulkarrivalat{t}>0}$, we have that
  $\BPGFof{\delay}{s} = \BPGFof{\queuesizepktleave}{ G_{\indicator{\bulkarrivalat{1}>0}}^{-1}\parantheses{s} }$, where
  \begin{IEEEeqnarray}{rCl}
  \BPGFof{\indicator{\bulkarrivalat{1}>0}}{s} &=& (1-\packetarrivalprob)^\framesize + s\parantheses{1-(1-\packetarrivalprob)^\framesize}. \label{expr:pgf_bulkarrival_greatherthanzero}
  \end{IEEEeqnarray}
  Furthermore, from the definition of $\arrivalsduringserving$ and from~\eqref{def:servicetime_bulk}, we obtain the equality
  \begin{IEEEeqnarray}{rCcCcl}
  \Bexpectation{\arrivalsduringserving} &=& \Bexpectation{\indicator{\bulkarrivalat{1}>0}}\Bexpectation{\servicetimeofpacket{1}}\Bexpectation{\bulkarrivalcountat{1}}
  &=&\packetarrivalprob\framesize\Bexpectation{\servicestoppingtime}.
  \end{IEEEeqnarray}
  We observe that
  $\BPGFof{\arrivalsduringserving}{s} =
    \BPGFof{\servicetimeofbulk{1}}{\BPGFof{\indicator{\bulkarrivalat{1}>0}}{s}}$,
    $\BPGFof{\servicetimeofpacket{1}}{s} = \BPGFof{\servicestoppingtime}{s}$, %
  and $\BPGFof{\servicetimeofbulk{1}}{s}
  = \BPGFof{\bulkarrivalcountat{1}}{\BPGFof{\servicetimeofpacket{1}}{s}}$ %
  where
  \begin{IEEEeqnarray}{rCl}
  \BPGFof{\bulkarrivalcountat{1}}{s} &=& \parantheses{\parantheses{1-\packetarrivalprob+\packetarrivalprob s}^\framesize-\parantheses{1-\packetarrivalprob}^\framesize}/\parantheses{1-\parantheses{1-\packetarrivalprob}^\framesize}.\nonumber
  \end{IEEEeqnarray}
  Algebraic manipulations yield~\eqref{expr:pgf_steadystate_delay}.%

\section{Computation of the CCDF from the PGF of a Random Variable}\label{appendix:CCDF_to_PGF}
We elaborate on the method used to compute the CCDF of a nonnegative integer-valued random variable~$\bX$ given its PGF~$\BPGFof{\bX}{s}$.
The chosen method is the recursion based $z$-transform inversion, which is exact, but it is applicable only when~$\BPGFof{\bX}{s}$ is a rational function.

We start by noting that the $z$-transform of the CCDF can be written as
\begin{IEEEeqnarray}{rCl}
\sum_{k=0}^\infty s^k \probof{\bX > k } &=& \frac{1-\BPGFof{\bX}{s}}{1-s}. \label{expr:ccdf_z_transform}
\end{IEEEeqnarray}
When~$\BPGFof{\bX}{s}$ is a rational function, we can rewrite~\eqref{expr:ccdf_z_transform} as
\begin{IEEEeqnarray}{rCl}
\sum_{k=0}^\infty s^k \probof{\bX > k } &=& \parantheses{\sum_{k=0}^n a_k s^k}/\parantheses{\sum_{k=0}^m b_k s^k} \label{expr:ccdf_z_transform_general}
\end{IEEEeqnarray}
for some positive integers $m,n$ and real numbers~$a_0,\dots,a_n,b_0,\dots,b_m$.
Multiplying by~$\sum_{k=0}^m b_k s^k$ both sides of~\eqref{expr:ccdf_z_transform_general} and equating the coefficients, we obtain
\[ \probof{\bX > k }  = \frac{1}{b_0}\parantheses{a_k - \sum_{u=1}^{k}b_{u} \probof{\bX > k-u } }\]
where~$a_k=0$ for~$k>n$ and~$b_u=0$ for~$u>m$.
The method is referred to as recursive because, in order to compute~$\probof{\bX > k }$, we need to compute first~$\probof{\bX > u }$ for all values of $u=0,\dots,k-1$.

\section{Proof of Theorem~\ref{thm:vlsf_achievability}}\label{proof:vlsf_achievability}
The choice of the auxiliary random variable $\commonrand$ (see Def.~\ref{def:vlsf_codes}), the random codebook construction, and the proof of~\eqref{expr:errorprob_bound} are the same as in the proof of \cite[Thm.~3]{polyanskiy2011feedback} and, hence, omitted.
To prove~\eqref{expr:stoppingtime_bound}, we first need to establish some notation.
Let $\codeword_1,\dots,\codeword_{2^\infobits}$ be the $2^k$ random codewords.
At time instant $\framesize t$, $t=1,2,\dots$ the decoder computes the $2^k$ information densities $\infodensity{\codeword^{\framesize t}_\Rinpmessage}{\chout^{\framesize t}}$, $\Rinpmessage \in \messagespace$ and feeds back an ACK as soon as one of the information densities exceeds~$\infodensitythreshold$.
Specifically, let~$\stoppingtime_{\Rinpmessage} = \inf\left\{ t\geq 0: \infodensity{\codeword_{\Rinpmessage}^{\framesize t}}{\chout^{\framesize t} }> \infodensitythreshold\right\}$.
The ACK is sent after frame~$    \stoppingtime = \min_{\Rinpmessage \in \messagespace} \stoppingtime_{\Rinpmessage}$
%
and the output of the decoder is the message with index $\max \{ \Rinpmessage: \stoppingtime_{\Rinpmessage}=\stoppingtime\}$.
The CCDF of $\stoppingtime$ can be upper-bounded as
\begin{IEEEeqnarray}{rCl}
\probof{\servicestoppingtime\geq t} &=& \frac{1}{2^\infobits}\sum_{\Rinpmessage\in\messagespace} \probof{\stoppingtime\geq t\given\inpmessage=\Rinpmessage} \\
&\leq& \frac{1}{2^\infobits}\sum_{\Rinpmessage\in\messagespace} \probof{\stoppingtime_\Rinpmessage\geq t\given\inpmessage=\Rinpmessage}\\
&=&  \probof{\thresholdstoppingtime\geq t}
\end{IEEEeqnarray}
where $\thresholdstoppingtime$ is defined in~\eqref{eq:info_dens_threshold}.
\section{Proof of Theorem~\ref{thm:peakage_ATL_ARQ_sync}}\label{proof:peakage_ATL_ARQ_sync}
\newtext{
Let $\arrivaltimeofbulk{}$ be the number of frames elapsed between the arrival of two consecutive packets, which is geometrically distributed.
 Then, $\BPGFof{\arrivaltimeofbulk{}}{s}$ takes the form given in~\eqref{expr:pgf_arrival_bulk}.
Note that under \ATLname, preemption occurs with probability $1-\packetdeliveredprob$, where $\packetdeliveredprob$ is given in~\eqref{expr:atl_pacekt_delivery_prob}.
Let~$\failedpacketscount$ be the number of packets  whose transmission is interrupted because of preemption, in between  two successfully delivered packets.
Then, given~$\failedpacketscount$, the peak age of information has the same distribution as the following sum of independent random variables: $\peakageofinfo \sim \arrivaltimeofbulk{}^{(0)} + \sum_{j=1}^{\failedpacketscount} \arrivaltimeofbulk{j}^{(1)} + \servicetimeofpacket{}^{(0)}$.
Here, $\arrivaltimeofbulk{}^{(0)}$ has the same distribution as $\arrivaltimeofbulk{}$   given that the first packet is delivered successfully,  whereas $\arrivaltimeofbulk{j}^{(1)}$ has the same distribution as $\arrivaltimeofbulk{}$  given that the first packet is not delivered because of preemption.
Furthermore,~$\servicetimeofpacket{}^{(0)}$ is the service time of a packet given that the packet is delivered (no preemption).
One can show that  $\arrivaltimeofbulk{j}^{(1)}\sim\servicetimeofpacket{}^{(0)}\sim\Bgeometricdist{1-\arqblockerrorprob\parantheses{1-\packetarrivalprob}^{\framesize}}$, which implies~\eqref{expr:ktn_ktl_steady_state_service_no_arrivals}, and also that $\BPGFof{\peakageofinfo}{s} = \Bexpectation{\BPGFof{\arrivaltimeofbulk{}^{(0)}}{s}\parantheses{\BPGFof{\servicetimeofpacket{}^{(0)}}{s}}^{\failedpacketscount+1}}$.
%
%
 To compute $\BPGFof{\arrivaltimeofbulk{}^{(0)}}{s}$, we use that
 $\BPGFof{\arrivaltimeofbulk{}}{s}=\packetdeliveredprob\BPGFof{\arrivaltimeofbulk{}^{(0)}}{s}+(1-\packetdeliveredprob)\BPGFof{\arrivaltimeofbulk{j}^{(1)}}{s}$
%
%
from which~\eqref{expr:atl_steady_state_service_one_arrivals} follows.
Finally, we obtain~\eqref{expr:peakage_ATL_ARQ_sync} by using that $\failedpacketscount + 1 \sim  \Bgeometricdist{\packetdeliveredprob}$}.

\end{appendices}

\linespread{0.99} 
\bibliographystyle{IEEEtran}
\bibliography{IEEEabrv,publishers,confs-jrnls,refs}

\end{document}